\documentclass{emulateapj}
\usepackage{amsmath}
\usepackage{epsfig}
\usepackage{epsf}
\input{epsf}
\usepackage{epstopdf}
\usepackage{amssymb}
\usepackage{bbold}
\usepackage{natbib}
\usepackage{graphicx}
\usepackage{color}
\usepackage{hyperref}
\usepackage{subfigure}
\usepackage{threeparttable}
\usepackage{xspace}
\usepackage{float}

\slugcomment{submitted to ApJ}

\DeclareGraphicsExtensions{.jpg,.pdf,.png,.eps,.ps}
\graphicspath{{FIGURES/}}

\newcommand{\Tcmb}{\mbox{$T_{\mbox{\tiny CMB}}$}}

\newcommand{\sqdeg}{\mbox{deg$^2$}}
\newcommand{\chisq}{\ensuremath{\chi^2}}
\newcommand{\delchisq}{\ensuremath{\Delta\chi^2}}

\newcommand{\ltsima}{$\; \buildrel < \over \sim \;$}
\newcommand{\ltsim}{\lower.5ex\hbox{\ltsima}}

\newcommand{\beq}{\begin{equation}}
\newcommand{\eeq}{\end{equation}}
\newcommand{\alphathresh}{1.5}
\newcommand{\um}{\ensuremath{\mu \mathrm{m}}}
\newcommand{\strue}{\ensuremath{S_\mathrm{true}}}
\newcommand{\smeas}{\ensuremath{S_\mathrm{meas}}}
\newcommand{\smax}{\ensuremath{S_\mathrm{max}}}
\newcommand{\smeasnine}{\ensuremath{S^\mathrm{meas}_{95}}}
\newcommand{\smeasonef}{\ensuremath{S^\mathrm{meas}_{150}}}
\newcommand{\smeastwot}{\ensuremath{S^\mathrm{meas}_{220}}}
\newcommand{\smaxnine}{\ensuremath{S^\mathrm{max}_{95}}}
\newcommand{\smaxonef}{\ensuremath{S^\mathrm{max}_{150}}}
\newcommand{\smaxtwot}{\ensuremath{S^\mathrm{max}_{220}}}

\newcommand{\alphatwosync}{\ensuremath{\alpha^{150,\mathrm{sync}}_{220}}}
\newcommand{\alphatwodust}{\ensuremath{\alpha^{150,\mathrm{dust}}_{220}}}
\newcommand{\alphao}{\alpha^{95}_{150}}
\newcommand{\alphaod}{\alpha^{95\mathrm{, dist}}_{150}}
\newcommand{\alphaor}{\alpha^{95\mathrm{, raw}}_{150}}
\newcommand{\alphat}{\alpha^{150}_{220}}
\newcommand{\alphatd}{\alpha^{150\mathrm{, dist}}_{220}}
\newcommand{\alphatr}{\alpha^{150\mathrm{, raw}}_{220}}

\def\Msun{M_\odot}
\def\Lsun{L_\odot}

\def \twthree {\textsc{ra23h30dec-55}}
\def \twonesix {\textsc{ra21hdec-60}}
\def \twonefif {\textsc{ra21hdec-50}}
\def \three {\textsc{ra3h30dec-60}}
\def \fiveh {\textsc{ra5h30dec-55}}

\def\KICPChicago{1}
\def\AAUChicago{2}
\def\Caltech{3}
\def\UChicago{4}
\def\ESO{5}
\def\Diego{6}
\def\Colorado{7}
\def\EFIChicago{8}
\def\Paris{9}
\def\PhysicsUChicago{10}
\def\Cavendish{11}
\def\Argonne{12}
\def\Dalhousie{13}
\def\Cambridge{14}
\def\NIST{15}
\def\McGill{16}
\def\Berkeley{17}
\def\Davis{18}
\def\LBNL{19}
\def\Arizona{20}
\def\Michigan{21}
\def\Munich{22}
\def\ExcellenceCluster{23}
\def\MPE{24}
\def\CaseWestern{25}
\def\Minnesota{26}
\def\STSI{27}
\def\ArtInstChicago{28}
\def\CfA{29}
\def\ITAToronto{30}
\def\Dunlap{31}
\def\AAToronto{32}

\begin{document}

\title{Extragalactic millimeter-wave point source catalog, number counts and statistics from 771 \sqdeg~of the SPT-SZ Survey}

\author{
  L.~M.~Mocanu,\altaffilmark{\KICPChicago,\AAUChicago}
  T.~M.~Crawford,\altaffilmark{\KICPChicago,\AAUChicago}
  J.~D.~Vieira,\altaffilmark{\Caltech}
  K.~A.~Aird,\altaffilmark{\UChicago}
  M.~Aravena,\altaffilmark{\ESO,\Diego}
  J.~E.~Austermann,\altaffilmark{\Colorado}
  B.~A.~Benson,\altaffilmark{\KICPChicago,\EFIChicago}
  M.~B\'{e}thermin,\altaffilmark{\Paris}
  L.~E.~Bleem,\altaffilmark{\KICPChicago,\PhysicsUChicago}
  M.~Bothwell,\altaffilmark{\Cavendish}
  J.~E.~Carlstrom,\altaffilmark{\KICPChicago,\AAUChicago,\EFIChicago,\PhysicsUChicago,\Argonne}
  C.~L.~Chang,\altaffilmark{\KICPChicago,\EFIChicago,\Argonne}
  S.~Chapman,\altaffilmark{\Dalhousie,\Cambridge}
  H-M.~Cho,\altaffilmark{\NIST}
  A.~T.~Crites,\altaffilmark{\KICPChicago,\AAUChicago}
  T.~de~Haan,\altaffilmark{\McGill}
  M.~A.~Dobbs,\altaffilmark{\McGill}
  W.~B.~Everett,\altaffilmark{\Colorado}
  E.~M.~George,\altaffilmark{\Berkeley}
  N.~W.~Halverson,\altaffilmark{\Colorado}
  N.~Harrington,\altaffilmark{\Berkeley}
  Y.~Hezaveh,\altaffilmark{\McGill}
  G.~P.~Holder,\altaffilmark{\McGill}
  W.~L.~Holzapfel,\altaffilmark{\Berkeley}
  S.~Hoover,\altaffilmark{\KICPChicago,\PhysicsUChicago}
  J.~D.~Hrubes,\altaffilmark{\UChicago}
  R.~Keisler,\altaffilmark{\KICPChicago,\PhysicsUChicago}
  L.~Knox,\altaffilmark{\Davis}
  A.~T.~Lee,\altaffilmark{\Berkeley,\LBNL}
  E.~M.~Leitch,\altaffilmark{\KICPChicago,\AAUChicago}
  M.~Lueker,\altaffilmark{\Caltech}
  D.~Luong-Van,\altaffilmark{\UChicago}
  D.~P.~Marrone,\altaffilmark{\Arizona}
  J.~J.~McMahon,\altaffilmark{\Michigan}
  J.~Mehl,\altaffilmark{\KICPChicago,\Argonne}
  S.~S.~Meyer,\altaffilmark{\KICPChicago,\AAUChicago,\EFIChicago,\PhysicsUChicago}
  J.~J.~Mohr,\altaffilmark{\Munich,\ExcellenceCluster,\MPE}
  T.~E.~Montroy,\altaffilmark{\CaseWestern}
  T.~Natoli,\altaffilmark{\KICPChicago,\PhysicsUChicago}
  S.~Padin,\altaffilmark{\KICPChicago,\AAUChicago,\Caltech}
  T.~Plagge,\altaffilmark{\KICPChicago,\AAUChicago}
  C.~Pryke,\altaffilmark{\Minnesota}
  A.~Rest,\altaffilmark{\STSI}
  C.~L.~Reichardt,\altaffilmark{\Berkeley}
  J.~E.~Ruhl,\altaffilmark{\CaseWestern}
  J.~T.~Sayre,\altaffilmark{\CaseWestern}
  K.~K.~Schaffer,\altaffilmark{\KICPChicago,\EFIChicago,\ArtInstChicago}
  E.~Shirokoff,\altaffilmark{\Berkeley}
  H.~G.~Spieler,\altaffilmark{\LBNL}
  J.~S.~Spilker,\altaffilmark{\Arizona}
  B.~Stalder,\altaffilmark{\CfA}
  Z.~Staniszewski,\altaffilmark{\CaseWestern}
  A.~A.~Stark,\altaffilmark{\CfA}
  K.~T.~Story,\altaffilmark{\KICPChicago,\PhysicsUChicago}
  E.~R.~Switzer,\altaffilmark{\ITAToronto}
  K.~Vanderlinde,\altaffilmark{\Dunlap,\AAToronto} and
  R.~Williamson\altaffilmark{\KICPChicago,\AAUChicago}
}

\altaffiltext{\KICPChicago}{Kavli Institute for Cosmological Physics, University of Chicago, Chicago, IL, USA 60637}
\altaffiltext{\AAUChicago}{Department of Astronomy and Astrophysics, University of Chicago, Chicago, IL, USA 60637}
\altaffiltext{\Caltech}{California Institute of Technology, Pasadena, CA, USA 91125}
\altaffiltext{\UChicago}{University of Chicago, Chicago, IL, USA 60637}
\altaffiltext{\ESO}{European Southern Observatory, Alonso de C\'{o}rdova 3107, Vitacura Santiago, Chile}
\altaffiltext{\Diego}{Universidad Diego Portales, Faculty of Engineering, Av. Ej\'{e}rcito 441, Santiago, Chile}
\altaffiltext{\Colorado}{Department of Astrophysical and Planetary Sciences and Department of Physics, University of Colorado, Boulder, CO, USA 80309}
\altaffiltext{\EFIChicago}{Enrico Fermi Institute, University of Chicago, Chicago, IL, USA 60637}
\altaffiltext{\Paris}{Laboratoire AIM-Paris-Saclay, CEA/DSM/Irfu - CNRS - Universit\'{e} Paris Diderot, CEA-Saclay, Orme des Merisiers, F-91191 Gif-sur-Yvette, France}
\altaffiltext{\PhysicsUChicago}{Department of Physics, University of Chicago, Chicago, IL, USA 60637}
\altaffiltext{\Cavendish}{Cavendish Laboratory, University of Cambridge, 19 J.J. Thomson Avenue, Cambridge, CB3 0HE, UK}
\altaffiltext{\Argonne}{Argonne National Laboratory, Argonne, IL, USA 60439}
\altaffiltext{\Dalhousie}{Department of Physics and Atmospheric Science, Dalhousie University, Halifax, NS B3H 3J5, Canada}
\altaffiltext{\Cambridge}{Institute of Astronomy, University of Cambridge, Madingley Road, Cambridge CB3 0HA, UK}
\altaffiltext{\NIST}{NIST Quantum Devices Group, Boulder, CO, USA 80305}
\altaffiltext{\McGill}{Department of Physics, McGill University, Montreal, Quebec H3A 2T8, Canada}
\altaffiltext{\Berkeley}{Department of Physics, University of California, Berkeley, CA, USA 94720}
\altaffiltext{\Davis}{Department of Physics, University of California, Davis, CA, USA 95616}
\altaffiltext{\LBNL}{Physics Division, Lawrence Berkeley National Laboratory, Berkeley, CA, USA 94720}
\altaffiltext{\Arizona}{Steward Observatory, University of Arizona, 933 North Cherry Avenue, Tucson, AZ, USA 85721}
\altaffiltext{\Michigan}{Department of Physics, University of Michigan, Ann  Arbor, MI, USA 48109}
\altaffiltext{\Munich}{Department of Physics, Ludwig-Maximilians-Universit\"{a}t, 81679 M\"{u}nchen, Germany}
\altaffiltext{\ExcellenceCluster}{Excellence Cluster Universe, 85748 Garching, Germany}
\altaffiltext{\MPE}{Max-Planck-Institut f\"{u}r extraterrestrische Physik, 85748 Garching, Germany}
\altaffiltext{\CaseWestern}{Physics Department, Center for Education and Research in Cosmology and Astrophysics, Case Western Reserve University,Cleveland, OH, USA 44106}
\altaffiltext{\Minnesota}{Department of Physics, University of Minnesota, Minneapolis, MN, USA 55455}
\altaffiltext{\STSI}{Space Telescope Science Institute, 3700 San Martin Dr., Baltimore, MD, USA 21218}
\altaffiltext{\ArtInstChicago}{Liberal Arts Department, School of the Art Institute of Chicago, Chicago, IL, USA 60603}
\altaffiltext{\CfA}{Harvard-Smithsonian Center for Astrophysics, Cambridge, MA, USA 02138}
\altaffiltext{\ITAToronto}{Canadian Institute for Theoretical Astrophysics, University of Toronto, 60 St. George St., Toronto, Ontario, M5S 3H8, Canada}
\altaffiltext{\Dunlap}{Dunlap Institute for Astronomy \& Astrophysics, University of Toronto, 50 St George St, Toronto, ON, M5S 3H4, Canada}
\altaffiltext{\AAToronto}{Department of Astronomy \& Astrophysics, University of Toronto, 50 St George St, Toronto, ON, M5S 3H4, Canada}

\email{lmocanu@uchicago.edu}
 
\begin{abstract}
We present a point source catalog from 771 \sqdeg~of the South Pole Telescope Sunyaev Zel'dovich (SPT-SZ) survey at 95, 150, and 220~GHz. We detect 1545 sources above 4.5$\sigma$ significance in at least one band. Based on their relative brightness between survey bands, we classify the sources into two populations, one dominated by synchrotron emission from active galactic nuclei, and one dominated by thermal emission from dust-enshrouded star-forming galaxies. We find 1238 synchrotron and 307 dusty sources. We cross-match all sources against external catalogs and find 189 unidentified synchrotron sources and 189 unidentified dusty sources. The dusty sources without counterparts are good candidates for high-redshift, strongly lensed submillimeter galaxies. We derive number counts for each population from 1 Jy down to roughly 9, 5, and 11 mJy at 95, 150, and 220~GHz. We compare these counts with galaxy population models and find that none of the models we consider for either population provide a good fit to the measured counts in all three bands. The disparities imply that these measurements will be an important input to the next generation of millimeter-wave extragalactic source population models.

\end{abstract}


\bigskip\bigskip


\section{Introduction}
\label{sec:intro}

Emission from extragalactic sources produces bright features on small angular scales in the millimeter-wavelength sky. 
These sources can be divided into two broad populations: sources with flux that is flat or decreasing with frequency, consistent with synchrotron emission from active galactic nuclei (AGN), and sources with flux increasing with frequency, consistent with thermal emission from dust-enshrouded star-forming galaxies (DSFGs). 
Synchrotron emission is produced by relativistic electrons in active galaxies; DSFGs emit light when photons from hot, young stars are absorbed by dust grains and reradiated at longer wavelengths.

The synchrotron-dominated population has been explored for decades through long-wavelength radio surveys. 
Many of these sources are strong emitters down to millimeter wavelengths. A thorough review can be found in \citet{dezotti10}. 
This population is generally split into ``steep-spectrum'' and ``flat-spectrum'' sources.

In the context of the unified AGN scheme, flat- and steep-spectrum sources are regarded as the same type of intrinsic object (an AGN-powered radio source), only observed at different orientations relative to the jet. 
When the line of sight is closely aligned with the relativistic jet, the source appears as a flat-spectrum blazar, showing compact, Doppler-boosted emission from the optically thick jet. 
The flat spectrum is believed to originate from the superposition of different self-absorbed components of the relativistic jets that have different self-absorption frequencies.
There are two main categories of blazars: BL Lacs and flat spectrum radio quasars (FSRQs), distinguished mainly by the fact that FSRQs exhibit strong emission lines.
In contrast, when the source is observed side-on, emission originates mainly in the extended, optically thin radio lobes. These sources are classified as steep-spectrum radio galaxies and are mostly associated with radio luminous elliptical and S0 galaxies.
Generally, surveys at 5 GHz and higher are dominated by flat-spectrum sources \citep{dezotti10}. In particular, FSRQ sources are expected to be the dominant source population at millimeter wavelengths above $\sim10$ mJy.

In recent surveys, blazars have been observed to exhibit a break in the synchrotron spectrum at frequencies around 100 GHz \citep{tucci11}. This steepening is caused by the transition from optically thick to optically thin emission from the jet due to energy losses of relativistic electrons through radiation (electron cooling).
Gigahertz-Peaked Spectrum Sources (GPS), with spectral peaks in the GHz range due to synchrotron self-absorption, have also been reported \citep{odea98}.
An essential characteristic of radio sources is their variability due to relativistic shocks in the jet; this can lead to biases in the spectral behavior assessed using non-simultaneous observations.

Owing to the recent expansion of millimeter-wave and submillimeter-wave observing capabilities, the dusty source population is now undergoing extensive characterization \citep[e.g.,][]{lagache05}.  
Statistically significant studies of this population began with the detection of the cosmic infrared background (CIB) in 1996 by the Far Infrared Absolute Spectrophotometer (FIRAS) on the Cosmic Background Explorer (COBE) satellite \citep{puget96}. 
The CIB is primarily comprised of the integrated light from DSFGs.
It was found that half of the energy emitted by galaxies over the history of the universe is in the CIB; optical and UV light is absorbed by dust and reradiated in the far-infrared \citep{dwek98, dole06}. 
The Infrared Astronomy Satellite (IRAS) carried out the first all-sky survey in the mid- and far-infrared (e.g. see review by \citet{sanders96}), detecting about 20,000 extragalactic sources, most of them at low redshift ($z<0.3$). 
These sources are now known as luminous and ultraluminous infrared galaxies (LIRGs and ULIRGs). While normal spiral galaxies have luminosities of roughly $10^{10}\rm\ \Lsun$ in the far-infrared, ULIRGs have over $10^{12}\rm\ \Lsun$ in this band. 
The total infrared luminosity of the LIRGs and ULIRGs makes up only a small fraction of the local infrared energy output \citep{soifer91};
however, these galaxies dominate the infrared emission at higher redshifts \citep{lefloch05}. 

The Submillimetre Common-User Bolometer Array (SCUBA) camera on the 15 meter James Clerk Maxwell Telescope (JCMT) at Mauna Kea in Hawaii was used to perform the first blank-field submillimeter survey to mJy depths at 850~\um\ \citep{smail97,hughes98c}, followed soon after by surveys to similar depths with millimeter-wave instruments such as the Max-Planck-Millimeter-Bolometer \citep[MAMBO,][]{greve04} at 1200~\um.
These surveys revealed a population of luminous, high-redshift DSFGs, which were coined as submillimeter galaxies or SMGs, as the bulk of their energy is emitted at submillimeter wavelengths \citep[see review by, e.g.,][]{blain02}.

Since SMGs were discovered, numerous studies have been undertaken to understand this source population and its properties \citep[e.g.,][]{ivison02, chapman05}. 
The spectral energy distribution (SED) of SMGs is well described by a modified blackbody spectrum at a temperature of roughly 30 K \citep{kovacs06,magnelli12}. They have stellar masses around $10^{11} \Msun$ and total infrared luminosities of $10^{13} \Lsun$. They are copiously forming stars at rates of $100-1000\ \rm\Msun/year$ and are most numerous at redshift $z$$\sim$2.5 \citep{chapman05} with a high redshift tail extending out to $z>6$ \citep{riechers13, weiss13, vieira13}. 
It is generally thought that the majority of bright SMGs originate in mergers \citep{engel10},
which cause their high star formation rates; this in turn leads to many supernova explosions and to the production of large quantities of dust \citep{gall11}.
SMGs are among the largest gravitationally bound objects at their epochs and are precursors to the most massive galaxies \citep{blain04}. 

A remarkable property of SMGs is that they can be detected from 500~\um\ to about 2 mm independently of redshift, such that the luminosity is roughly proportional to the flux for $1<z<10$.  This is due to the fact that redshift dimming is compensated by observing the galaxy closer to the peak of its spectral energy distribution \citep[negative K-correction,][]{blain96}. Also, this implies that measurements of the CIB at 220 GHz (1.4 mm) are sensitive to the complete history of emission from DSFGs. 

\citet[hereafter V10]{vieira10} reported the discovery of a population of very bright and rare dust-dominated sources in $\sim$100 \sqdeg\ of South Pole Telescope (SPT) data. These sources had no counterparts in the IRAS catalog, implying that they could not be members of the local U/LIRG population. Consequently, they were hypothesized to be either local galaxies with dust temperatures too cold to be detected by IRAS, or high-redshift galaxies, either intrinsically ultra-bright, or strongly lensed by massive galaxies or clusters along the line of sight---as lensing increases the observed flux, making the sources appear brighter. Theoretical models had previously predicted such a sample of strongly lensed SMGs \citep{blain96, negrello07}.

Subsequent follow-up of those sources confirmed that they are high-redshift, strongly lensed SMGs. The first line of evidence, outlined in \citet{greve12}, is based on Atacama Pathfinder Experiment (APEX) 850~\um\ and 350~\um\ follow-up on 11 of the brightest lensed candidates.  
The analysis  to determine the photometric redshifts of the sources in a statistical fashion found that these galaxies lie at a median redshift of $z \sim 3.3$, higher than previously identified SMG samples \citep[e.g.,][]{chapman05}, which, together with their observed flux, implies very high luminosities. However, compared to the empirical luminosity-temperature relation of the population of unlensed sources, their dust temperatures are characteristic of regular SMGs, arguing that these objects are unlikely to be so intrinsically luminous.
This suggests that the objects are strongly lensed members of the normal SMG population. Recently obtained Atacama Large Millimeter/submillimeter Array (ALMA) imaging and spectroscopy \citep{vieira13} of a larger ($N \sim 25$) sample of SPT-discovered sources selected from the catalog presented in this paper demonstrates  that they are indeed high-redshift objects---with a measured spectroscopic redshift distribution with a mean of $\bar{z}=3.5$ \citep{weiss13}---that are strongly lensed by foreground galaxies, with
most sources resolved into arcs or Einstein rings \citep{hezaveh13}.

The Herschel Multi-tiered Extragalactic Survey \citep[HerMES,][]{oliver10} and the Herschel Astrophysical Terahertz Large Area Survey \citep[H-ATLAS,][]{clements10} have also identified a population of very bright dusty sources.  
A  discussion of the detection of lensed SMGs based on Herschel data can be found in \citet{negrello10} and \citet{wardlow13}. Millimeter-wave point source catalogs and number counts have also recently been released by the Atacama Cosmology Telescope \citep[ACT,][]{marsden13} and Planck \citep{planck13-28}. 


The South Pole Telescope (SPT) has now completed a 2500 \sqdeg, three-band survey of the millimeter-wave sky. 
Due to the sensitivity and angular resolution of the SPT, this survey data contains a large number of extragalactic point sources (which are unresolved by the arcminute beam). These sources are of high astrophysical and cosmological interest, relevant for studying the early stages of galaxy formation and their subsequent evolution. The multi-band data allow differentiation between source populations. Apart from their astrophysical importance, emissive sources are also significant contaminants to the small-scale ($\ell \gtrsim 2500$) cosmic microwave background (CMB) power spectrum. 
Understanding the properties of these source populations is thus essential for CMB analyses, for instance for separating primary CMB anisotropy power from secondary effects such as lensing and Sunyaev-Zel'dovich (SZ) effects. 
Measurements of source counts also help constrain the point-source contribution to noise and bias in SZ galaxy cluster surveys. 

This is the second point source catalog paper released from the SPT-SZ survey. The previous point source analysis, V10, presented a point source catalog and number counts derived from an 87 \sqdeg~field surveyed by SPT in 2008, using only two-band data. The spectral index between 150 and 220 GHz was used to classify sources as synchrotron- or dust-dominated. This analysis improves upon previous results by bringing the total catalogued area to $\sim$771 \sqdeg~and extending the analysis to include the 95 GHz band.

This paper is organized as follows. 
SPT observations and the data reduction procedure are described in \S\ref{sec:observ}. The mapmaking and source-finding algorithms are also detailed in this section. 
In \S\ref{sec:deboost}, we present the flux deboosting procedure used to estimate the intrinsic fluxes and spectral indices of sources, and detail their classification as synchrotron-dominated or dust-dominated.
The source catalog and a discussion of extended sources and cross-matching with external catalogs are found in \S\ref{sec:catalog}.
Total and by-population source number counts in each observation band are presented in \S\ref{sec:number_counts}.
In \S\ref{sec:pop}, we discuss the source populations and compare the number counts to predictions of galaxy evolution models. We present conclusions in \S\ref{sec:concl}.


\section{Observations and data reduction} \label{sec:observ}

\subsection{Instrument and survey}

The South Pole Telescope (SPT) is a 10-meter telescope located at the Amundsen-Scott South Pole station in Antarctica \citep{carlstrom11}. At 150~GHz (2~mm), the SPT has arcminute angular resolution and a one square degree diffraction-limited field of view. The SPT was designed for high-sensitivity millimeter/sub-millimeter observations of faint, low-contrast sources, such as CMB anisotropies. The first survey with the SPT, designated as the SPT-SZ survey, was completed in November 2011 and covers a $\sim 2500$ deg$^2$ region of the southern extragalactic sky in three frequency bands, $95$, $150$, and $220$~GHz, corresponding to wavelengths of 3.2, 2.0, and 1.4~mm. The fields were surveyed to depths of approximately 40, 18, and 70\,$\mu{\rm K}$-arcmin at 95, 150, and 220 GHz respectively.\footnote{Throughout this work, the unit K refers to equivalent fluctuations in the CMB temperature, i.e., the temperature fluctuation of a 2.73 K blackbody that would be required to produce the same power fluctuation.  The conversion factor is given by the derivative of the blackbody spectrum $\frac{dB_{\nu}}{dT}$, evaluated at 2.73 K.}

\subsection{Observations}

This paper uses data from five fields observed by the SPT in 2008 and 2009. The fields are referred to using the J2000 coordinates of their centers, right ascension (RA) in hours and declination (DEC) in degrees. 
Table~\ref{tab:fields} lists the positions and effective areas of these fields.
These are the same fields used for the CMB power spectrum analysis in \citet{keisler11}.
The total effective area used for the catalog and analysis in this work is 771 \sqdeg. We use the previously released catalog exactly as it was analyzed in V10 and add 684~\sqdeg~of newly analyzed data.

The SPT-SZ camera focal plane was composed of six detector modules, each of which could be configured to observe in a different frequency band. In 2008, when the \fiveh~ and \twthree~fields were observed,
there were three modules operating at 150~GHz, two at 220~GHz, and one at 95~GHz; however, the 2008 95~GHz module did not produce survey-quality data. In 2009, when the \twonesix, \three, and \twonefif~fields were observed, one
of the 220~GHz modules was replaced by a fourth 150~GHz module, and the 95~GHz module was upgraded.
As a result, the depth of the fields in the three observing bands is different for the 2008 and 2009 observing seasons. In particular, the 
part of the catalog 
in this work that comes from the 2008 fields (\fiveh\ and \twthree) is derived from deeper 220 GHz data but has no 95~GHz data. As a result, the source selection (and relative contributions of the source populations) differs slightly from the catalog derived from the 2009 fields. However, the 150 GHz depths are similar for the two observing seasons.


\begin{deluxetable*}{ l c c c c c c c}
\tablecaption{SPT fields used in this work}
\tablehead{
\colhead{Name} & \colhead{Season} & \colhead{R.A. ($^\circ$)} & \colhead{Decl. ($^\circ$)} & \colhead{$\Delta$R.A. ($^\circ$)} & \colhead{$\Delta$Decl. ($^\circ$)} & \colhead{No. sectors} & \colhead{Effective Area (deg$^2$)} } 
\startdata
\fiveh        & 2008 & 82.5 & -55.0 & 15 & 10 & 3$\times$3 & 86.7 \\
\twthree      & 2008 & 352.5 & -55.0 & 15 & 10 & 3$\times$3 & 100.5 \\
\twonesix     & 2009 & 315.0 & -60.0 & 30 & 10 & 6$\times$3 & 153.5 \\
\three        & 2009 & 52.5  & -60.0 & 45 & 10 & 8$\times$3 & 232.0 \\
\twonefif     & 2009 & 315.0 & -50.0 & 30 & 10 & 6$\times$3 & 198.5 \\
\hline\\

Total         &       &       &    &    &  & & 771.2
\enddata

\label{tab:fields}
\tablecomments{ The locations and sizes of the fields included in this work.  For each field we give the center of the field in Right Ascension (R.A.) and Declination (Decl.), the extent of the field in Right Ascension and Declination, the number of sectors the field is divided into (see \S\ref{sec:matched}) and the effective field area as defined by the apodization mask.}
\end{deluxetable*}

SPT observations are performed by sequentially scanning across each field back and forth once at constant elevation, then taking a step up in elevation. One of the 2008 fields, \twthree, and the three fields from 2009, \twonesix, \three, and \twonefif, were observed using a lead-trail scan strategy, such that each field is divided into two halves in RA. The lead half is observed first. The trail half is then observed such that, due to the Earth's rotation, both are scanned at the same range of azimuth angle. This allows for removal of potential ground-synchronous signal; however, such a signal was not detected. Therefore, we coadd the lead and trail observations together into a single map. Additionally, about two thirds of the \twonefif~observations were performed using elevation scans. In this observing mode, the telescope scanned up and down in elevation (at roughly the same speed as in the azimuth scans) but did not move in azimuth, letting the sky field drift through the field of view. 

An observation, defined as a complete set of scans covering the field, takes from 30 minutes to a few hours, depending on the field being observed. The final maps for each field used in this work are made from 400 to 700 full observations of the field.


\subsection{Data reduction} 
\label{sec:reduction}

The data reduction procedure is described in detail in \citet{schaffer11}. We summarize the method and the differences from that analysis here and refer the reader to the paper for more details.

\subsubsection{Timestream filtering}
\label{sec:filtering}

Each detector measures the brightness temperature of the sky as a function of time. 
The time-ordered data (TOD) from well-performing detectors are grouped into scans, keeping only data from the regions observed with constant scan velocity. The TOD are recorded at 100~Hz, then filtered in the Fourier domain. In order to avoid noise aliasing, a 25~Hz low-pass filter is applied to each scan to remove signal on scales smaller than roughly 0.5 arcminutes---the Nyquist frequency corresponding to the pixel size of the final maps (0.25 by 0.25 arcmin). 

Fluctuations in atmospheric emission due to turbulent water vapor become important on large spatial scales, causing low-frequency noise in the TOD. Additionally, the readout system introduces ``$1/f$'' noise into the TOD. This low-frequency noise is mitigated by a two-step procedure. 

First, a Legendre polynomial (of first order for the azimuth scans and ninth order for the elevation scans) is subtracted from the TOD of each detector. 
Then, the TOD are high-pass filtered in the Fourier domain with a filter cutoff frequency corresponding to a spatial scale of 45 arcmin in the scan direction.

The atmospheric fluctuation signal is highly correlated between the detectors because the detector beams overlap in the turbulent layers of the atmosphere. 
For this reason, the average of all well-performing detectors in each module is removed from the TOD at each time step. This acts like an isotropic spatial high-pass filter with an angular scale of about 0.5 degrees.

The TOD filtering described here has the effect of altering the shape of point sources in the maps. 
In the absence of filtering, the shape of point sources in the maps would simply be the instrument point-spread function or beam. 
The high-pass filtering causes a ringing pattern around sources in the maps, particularly in the scan direction.
Moreover, the effects of filtering are map-position-dependent. Those effects are dealt with as described in \S\ref{sec:clean}.

\subsubsection{Mapmaking}

The next step is going from the TOD to maps of the sky. 
The pointing model has been described in \citet{schaffer11}. 
We approximate the sky as flat across each field, and use the oblique Lambert equal-area projection with 0.25 arcmin pixels. 
This projection preserves distances and areas across the field, such that the beam shape will not be distorted across the map, which is important for finding sources with CLEAN algorithm (described in \S\ref{sec:clean}). 
However, in this projection, the angle between the scan direction and the map rows varies with map position (see \S\ref{sec:matched}). 

Single-observation maps are made by averaging all TOD that fall in each pixel by inverse-variance weighting based on the detector power spectral densities between 1 - 3~Hz. 
Single-observation maps with exceedingly high noise are discarded.  
All maps that pass the cut are then coadded into a final map for each observing band.

The maps are calibrated as follows. 
The relative calibrations of the TOD between single observations are determined from measurements of the galactic HII region RCW38. 
The absolute calibration is obtained by comparing the SPT power spectrum for each season to the Wilkinson Microwave Anisotropy Probe \citep[WMAP7,][]{larson11} power spectrum across the multipole range $650 < \ell < 1000$. 
The uncertainty of this calibration in temperature is estimated to be 1.8\%, 1.6\% and 2.4\% at 95, 150, and 220~GHz \citep{reichardt12b}. 
These uncertainties are highly correlated because the main sources of error, WMAP7 bandpower errors and SPT sample variance, are nearly identical between bands. 
We set this band-to-band correlation factor to 1 in the uncertainty calculation.

The absolute pointing is calculated by comparing the locations of the brightest sources in each field to their coordinates in the Australia Telescope 20GHz (AT20G) Survey catalog \citep{murphy10}, which has 1\arcsec~RMS positional accuracy. The RMS positional uncertainty of the brightest $\sim$40 sources in each field after applying the pointing correction is roughly 4\arcsec~in declination and 4\arcsec~in cross-declination (defined as RA$\cdot\cos(\rm{Dec})$).


\subsection{Source-finding}

\subsubsection{Matched filter}
\label{sec:matched}
 
We construct a matched filter \citep{tegmark98} $\psi$ and apply it to the map in the Fourier domain to enhance the signal-to-noise of pointlike objects. 
The matched filter maximizes sensitivity to beam-sized features by downweighting larger and smaller angular scales where the noise is larger and/or signal is lower. 
\begin{equation}
\label{eqn:optfilt}
\psi \equiv \frac{\tau^T N^{-1}}{\sqrt{\tau^T N^{-1} \ \tau}}
\end{equation}
where $\tau$ is the source shape and $N$ is the noise covariance matrix. 
The precise source shape is determined by the convolution of the beam with the map-domain equivalent of all TOD filtering applied before mapmaking. 
Given that the TOD filtering (and thus source shape) is map-position-dependent, we divide each map into 3x3 (for \twthree), 6x3 (for \twonesix~and \twonefif), or 8x3 (for \three) sectors (as listed in Table~\ref{tab:fields}) and evaluate $\tau$ and the noise separately for each sector. 
To check whether these sector sizes are appropriate, we tested the effects of applying the mid-sector transfer function to sources at one side of the sector, and found those effects to be subdominant to the beam and calibration error even for the brightest sources, and generally a 1-2\% level effect for most sources.

The first ingredient needed for the filter is the beam shape. The SPT beams are measured using a combination of maps of Venus, Jupiter, and the brightest point sources in the fields. 
The main lobes are well approximated by Gaussian functions with FWHM of 1.7$'$, 1.2$'$, and 1.0$'$ (at 95, 150, and 220~GHz, respectively). 
Beam sidelobes are unimportant for the scales relevant to point source analysis, as they are filtered out.

The source shape is determined by constructing maps of simulated point sources in the following way. 
First, we place a delta function convolved with the beam at the center of each sector. 
We ``reobserve'' this signal using the real pointing information and the same TOD filtering as is applied to the real data. 
The result is a real-space representation of the source shape for each sector. 
By transforming this into the Fourier domain, we obtain two dimensional transfer functions (TF), representing the relative suppression of signal power due to the PSF and filtering as a function of angular scale along the map $x$ and $y$ directions. 

Map noise is comprised of instrumental and atmospheric noise and contributions from real astrophysical signal---namely primary and secondary CMB anisotropies (such as the SZ effect) and point sources below the detection threshold. The instrumental and atmospheric noise components are estimated using a jackknife technique. We take all single-observation maps for each band, multiply half of them by -1, and coadd them in order to remove all astrophysical signal. We call those maps ``difference maps''. This procedure is repeated many times, randomly dividing the single observations in half each time. The Fourier transforms of all difference maps are quadrature-averaged to obtain the two-dimensional noise power spectral density (PSD), which is equivalent to the noise covariance. An estimate of primary CMB anisotropy is then added to the noise covariance. For this, we take the standard $\Lambda$CDM model CMB power spectrum best-fit by WMAP7 \citep{larson11} and SPT data, as presented in \citet{keisler11}.
Contributions from secondary anisotropies and sources below the confusion limit are small and can be neglected when constructing the matched filter.

In summary, we use the TFs and noise PSDs to construct the matched filters $\psi$ to apply to each map sector.

\subsubsection{CLEAN procedure}
\label{sec:clean}

In the filtered maps, sources are located using the CLEAN algorithm \citep{hogbom74}.  
This algorithm was developed for producing maps in radio interferometry, where irregular baseline coverage or the finite number of antennae results in finite sampling of the Fourier plane. 
This incomplete mode sampling leads to a beam exhibiting sidelobes (``dirty beam"), which renders the resulting map difficult to interpret. 
We have a similar ``dirty beam", due in our case to the TOD filtering described in \ref{sec:filtering}. 

For each sector, we construct a source template $\tau^\prime$ by taking the source shape $\tau$, discussed in the previous section, and convolving it with the matched filter $\psi$: 

\begin{equation}
\label{eqn:dirtybeam}
\tau^\prime = \psi\tau.
\end{equation}

The CLEAN procedure is implemented as follows: 
\begin{itemize}
\item{Search for the brightest pixel in the map.}
\item{Construct a source template at the position of this brightest pixel by rotating the template $\tau$ at the center of the sector by the difference in angle between the scan direction at the position of the source and the scan direction at the center of the sector.}
\item{Subtract the filtered source template $\tau^\prime$ multiplied by a loop gain factor at the position of the peak. The loop gain is set to 0.1 to account for imperfect source templates and the presence of extended sources.} 
\item{Look for the brightest pixel in the resulting map and loop through this procedure until no peaks are left above the chosen detection threshold.}
\end{itemize}
We choose to run the source-finder down to a 4.5$\sigma$ level; this is the significance threshold of the final catalog. We chose this value as the threshold where the V10 catalog was found to be roughly 90\% pure. We denote the map that remains after performing all the subtractions as the residual map. All the brightest pixels detected by the algorithm are sorted by intensity and grouped into sources using a brightness-dependent association radius between 30 arcseconds and 2 arcminutes. The position of each source is taken to be the center of brightness of all pixels associated with the source.

The flux of each source is determined by taking the value of the brightest pixel corresponding to the source from the filtered map and converting it from CMB fluctuation temperature to units of flux, namely:
\beq
\label{eqn:temp2flux}
S[\mathrm{Jy}] = T_\mathrm{peak} \cdot \Delta \Omega_\mathrm{f} \cdot 10^{26} \cdot \frac{2k_{\rm B}}{c^2} \left ( \frac{k_{\rm B} \Tcmb}{h} \right )^2 \frac{x^4 e^x}{(e^x-1)^2},
\eeq
where $x=h\nu/(k_{\rm B} \Tcmb)$ and the effective solid angle under the source template $\Delta \Omega_\mathrm{f}$ is calculated from:
\beq
\label{eqn:area_eff}
\Delta \Omega_\mathrm{f} = \left [ \int d^2k \ \psi(k_x,k_y) \ \tau(k_x,k_y) \right ]^{-1},
\eeq
where $k_x$ and $k_y$ are the angular wavenumbers associated with the $x$ and $y$ coordinates of the map.

The residual map is visually inspected to check for the effectiveness of the procedure and to identify any extended sources. After visual inspection, we remove a few obviously spurious detections caused by CLEAN residuals near the brightest sources. These are consistent with the beam uncertainty. We also remove detections generated by the sidelobe response to extended sources.

The single-band catalogs are combined based on position offset between bands: sources are considered detected in more than one band if the distance between detections in different bands is less than 30 arcseconds. This 
radius is chosen as a compromise between falsely associating sources which are in fact independent detections
and missing true associations due to positional uncertainty. Thirty arcseconds is roughly 1.5 times the positional
uncertainty for a $4.5\sigma$ detection in the band with the widest beam (95~GHz).
We define the detection band of each source as the band in which the source is detected at the highest signal-to-noise ratio. The coordinates recorded in the catalog reflect the position of the source in the detection band. If a source is not detected in a band above the CLEAN cut-off significance, the flux in that band is taken to be the value of the pixel in the residual map at the location found in the detection band map. 



\section{Flux deboosting and corrected spectral indices} 
\label{sec:deboost}

The differential number counts, $dN/dS$, where $N$ is the number of sources with flux $S$, are expected to be a very steep function of flux, which leads to a positive bias in the measured fluxes. 
We refer to this effect as flux boosting. 
Effectively, it is more likely that a source of measured flux $S$ is intrinsically dimmer and standing on top of a positive noise fluctuation, rather than brighter and on top of a negative noise fluctuation. 
This occurs because, although Gaussian noise is equally likely to have a positive or negative contribution to the measured flux of a given source, there exist many more intrinsically dim sources. 
This bias is more pronounced for low signal-to-noise detections, and is closely related to what is referred to as ``Eddington bias'' \citep{teerikorpi04}. 
We note, however, that this latter term is generally used in the literature to describe the bias in estimating source counts as a function of brightness, as opposed to the brightness of individual sources.

There will also be a small positive flux bias due to selecting peaks in the map---or, equivalently, 
maximizing the signal over $x$ and $y$ \citep[e.g.,][]{austermann10}---and 
a small negative flux bias when taking the flux of a source
detected in one band from the residual map of a different band (due to positional uncertainty in the 
detection band). 
The relation of the apparent source signal-to-noise to the true signal-to-noise
due to maximizing over two parameters
is expected to be $S/N_\mathrm{app} = \sqrt{(S/N_\mathrm{true})^2 + 2}$ \citep[e.g.,][]{vanderlinde10}, which
is a $<5 \%$ effect at the $4.5\sigma$ threshold of this catalog and negligible at higher significances. 
The bias due to positional uncertainty is also expected to be very small in this catalog, because the 
positional uncertainty on $>4.5\sigma$ sources is a small fraction of the beam.

\subsection{Motivation for multi-band deboosting}
\label{sec:motiv}

\citet{crawford10} present a method for estimating the flux of individual sources from multifrequency survey data.
In what follows, we motivate this procedure and summarize its main steps. To correct a single-band flux measurement, the simplest attempt at a Bayesian approach would be to calculate
\begin{align}
\label{eq:bay4}
P(\strue|\smeas) \propto P(\smeas|\strue)P(\strue),
\end{align}
where $P(\strue|\smeas)$ is the posterior probability that the flux of a source is $\strue$ given a measured flux $\smeas$, $P(\smeas|\strue)$ is the likelihood of measuring a flux $\smeas$ given a flux $\strue$ (which in the simplest case is a Gaussian centered at $\strue$ with a width related to instrumental and atmospheric noise in the maps), and $P(\strue)$ is the prior probability of a source to have an intrinsic flux $\strue$ (which is proportional to the differential number counts $dN/dS$).

The first issue with applying the standard procedure separately to fluxes measured in three bands is that the flux priors are correlated between bands and cannot be directly separated into a product of one-dimensional distributions.

The second problem with this approach is that the measured flux in one pixel does not correspond to the flux of a single source, because fainter sources also contribute to the signal. Instead, it is more appropriate to look for the probability that the brightest source in a pixel has a true flux $\smax$, given that the total flux in the pixel was measured to be $\smeas$:
\begin{align}
\label{eq:bay2}
P(\smax|\smeas) \propto P(\smeas|\smax)P(\smax),
 \end{align}
where $P(\smeas|\smax)$ is the likelihood of measuring a total flux $\smeas$ in a pixel given that the brightest source in the pixel has a flux $\smax$, and $P(\smax)$ is the prior probability that the brightest source in the pixel has flux $\smax$.

Again, $P(\smeas|\smax)$ can be approximated by a Gaussian distribution which includes contributions from both faint sources and noise. 
This is because a large number of sources below the confusion limit contribute to the flux in a pixel, and thus the distribution of pixel fluxes approaches a Gaussian, as does the contribution from instrumental and atmospheric noise.

The prior $P(\smax)$ can be written as the probability that a source of flux $\smax$ exists in the pixel multiplied by the probability that no sources brighter than $\smax$ exist in the pixel and is proportional to the differential number counts $dN/dS$, but with an extra exponential suppression given by the mean number of sources with flux above $\smax$.

\citet{crawford10} developed a method to overcome this limitation and estimate individual source properties for the two-band case. This method was used to correct the source fluxes in V10. Here, we extend this calculation by adding a third band. 

\subsection{Method for simultaneous 3-band deboosting}
\label{sec:three_band}

Let $S_{95}$, $S_{150}$, and $S_{220}$ be the fluxes measured for a source in the $95$~GHz, $150$~GHz, and $220$~GHz bands respectively, and $\nu_{95}$, $\nu_{150}$, and $\nu_{220}$ be the effective band centers. 
For each source, we define two distinct spectral indices, $\alphao$ and $\alphat$, as the slope of the assumed power law behavior of the flux as a function of frequency between $95$~GHz~-~$150$~GHz and $150$~GHz~-~$220$~GHz, respectively:
\begin{eqnarray}
S_{95} &=& S_{150}\left(\frac{\nu_{95}}{\nu_{150}}\right)^{\alphao} \nonumber\\
S_{220} &=& S_{150}\left(\frac{\nu_{220}}{\nu_{150}}\right)^{\alphat}.
\label{eqn:def}
\end{eqnarray}
The effective band centers depend slightly on the spectral index of the source. We calculate the band centers of the SPT bands by assuming a spectral index of 0, which yields 97.6, 152.9, and 218.1 GHz. This approximation does not significantly affect the source fluxes reported here.
We want to obtain a three dimensional posterior probability density $P(\smaxnine,\smaxonef,\smaxtwot|\smeasnine,\smeasonef,\smeastwot)$ for the true values of the fluxes of the brightest source in a certain pixel in each band, given the measured fluxes. This can be expressed as:

\begin{eqnarray}
\label{eq:bay3}
P(\smaxnine,\smaxonef,\smaxtwot|\smeasnine,\smeasonef,\smeastwot) &\propto& \nonumber\\ 
P(\smeasnine,\smeasonef,\smeastwot|\smaxnine,\smaxonef,\smaxtwot)&\cdot& \nonumber\\
P(\smaxnine,\smaxonef,\smaxtwot).
\end{eqnarray}

Using a Gaussian likelihood approximation, we first calculate the likelihood $P(\smeasnine,\smeasonef,\smeastwot|\smaxnine,\smaxonef,\smaxtwot)$ to measure the fluxes $(\smeasnine,\smeasonef,\smeastwot)$, given that the true fluxes of the brightest source in the pixel are $(\smaxnine,\smaxonef,\smaxtwot)$:

\begin{eqnarray}
P(\smeasnine,\smeasonef,\smeastwot|\smaxnine,\smaxonef,\smaxtwot)&=& \nonumber\\
\frac{\exp\left(-\frac{1}{2}{\bf r}^{T}{\bf C}^{-1}{\bf r}\right)}{2\pi \sqrt{\rm{det\ {\bf C}}}}.
\label{eqn:smmeas}
\end{eqnarray}
Here, ${\bf C}$ is the noise covariance between bands. This includes contributions from the RMS of the coadded map for each band, beam calibration (both diagonal) and WMAP power calibration. Also, ${\bf r}$ is a residual vector defined as

\begin{eqnarray}
{\bf r} = \big\{\smeasnine-\smaxnine, \smeasonef-\smaxonef, \smeastwot-\smaxtwot\big\}. 
\label{eqn:arr}
\end{eqnarray}
%

\begin{figure*}[!ht] 
\begin{center} 
\subfigure{\includegraphics[width=8.9cm]{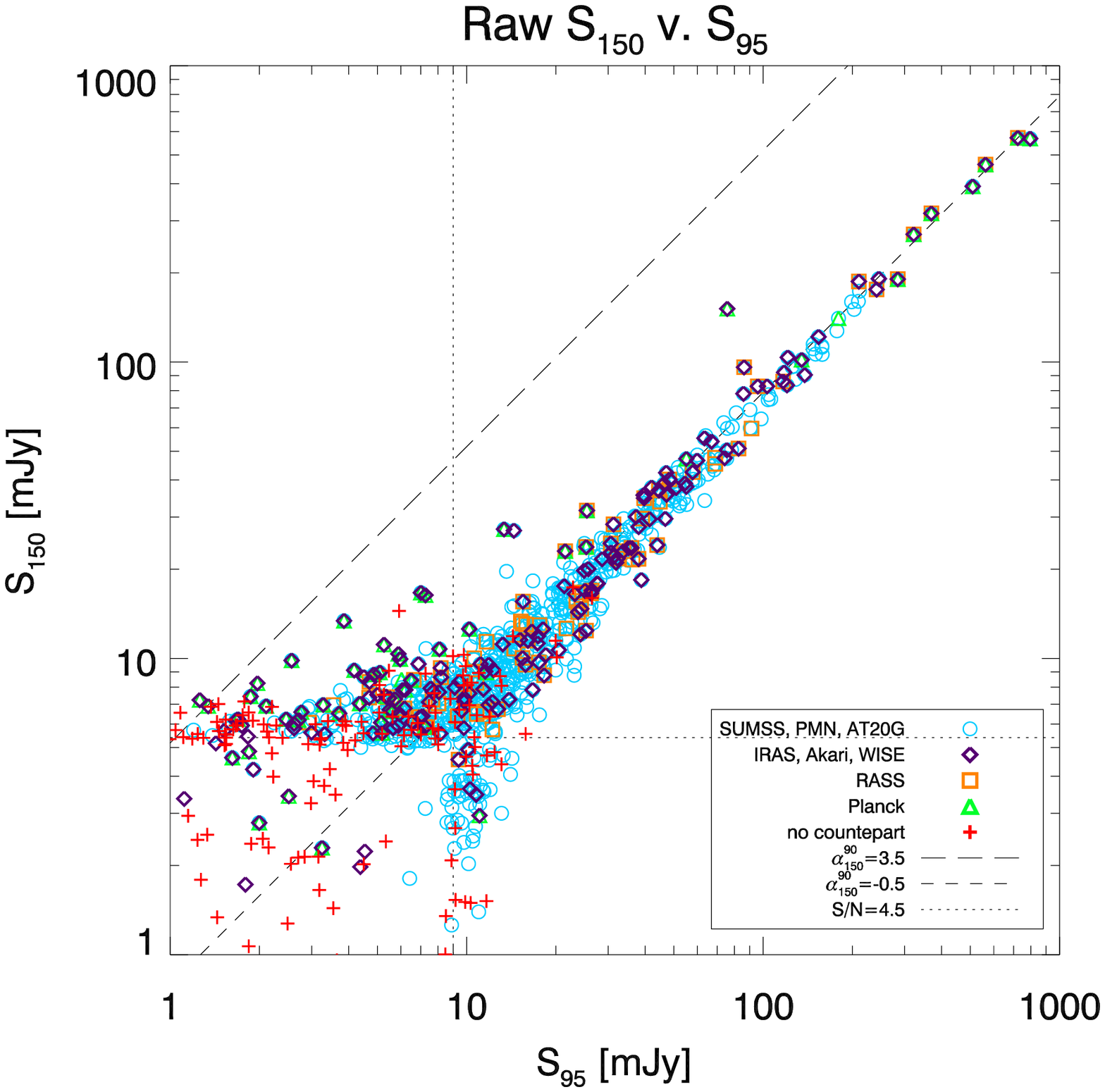}}
\subfigure{\includegraphics[width=8.9cm]{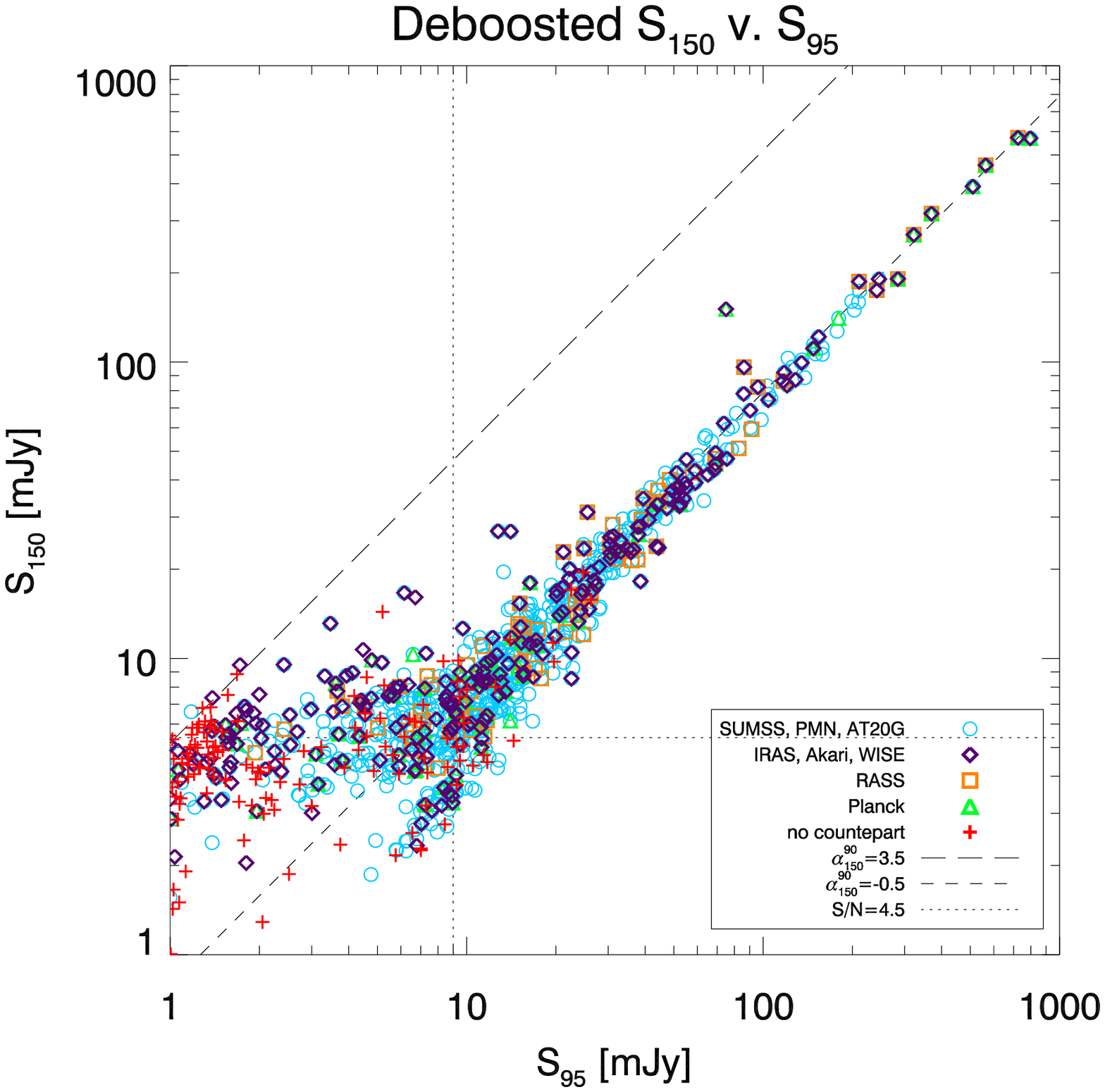}}
\subfigure{\includegraphics[width=8.9cm]{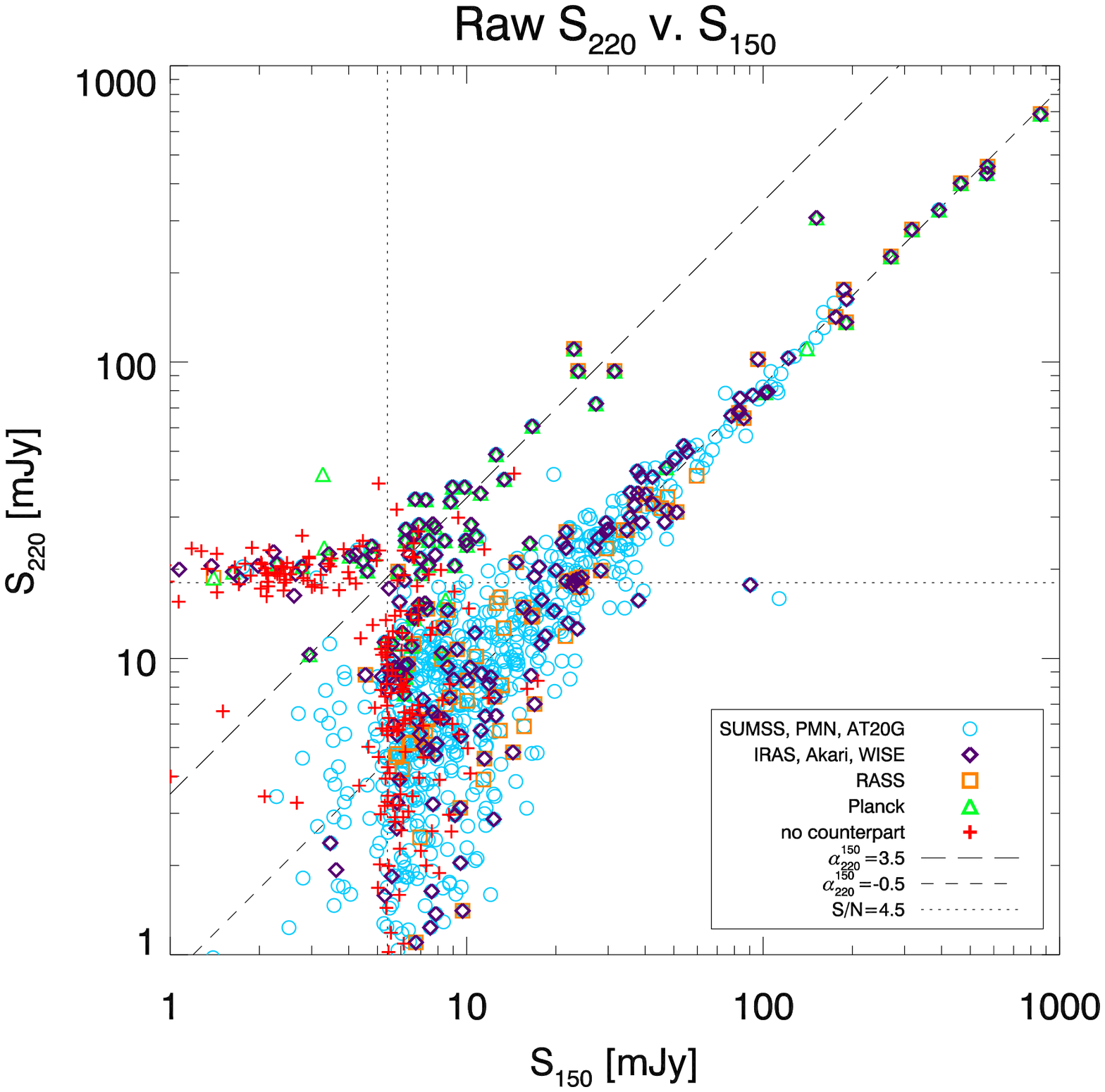}}
\subfigure{\includegraphics[width=8.9cm]{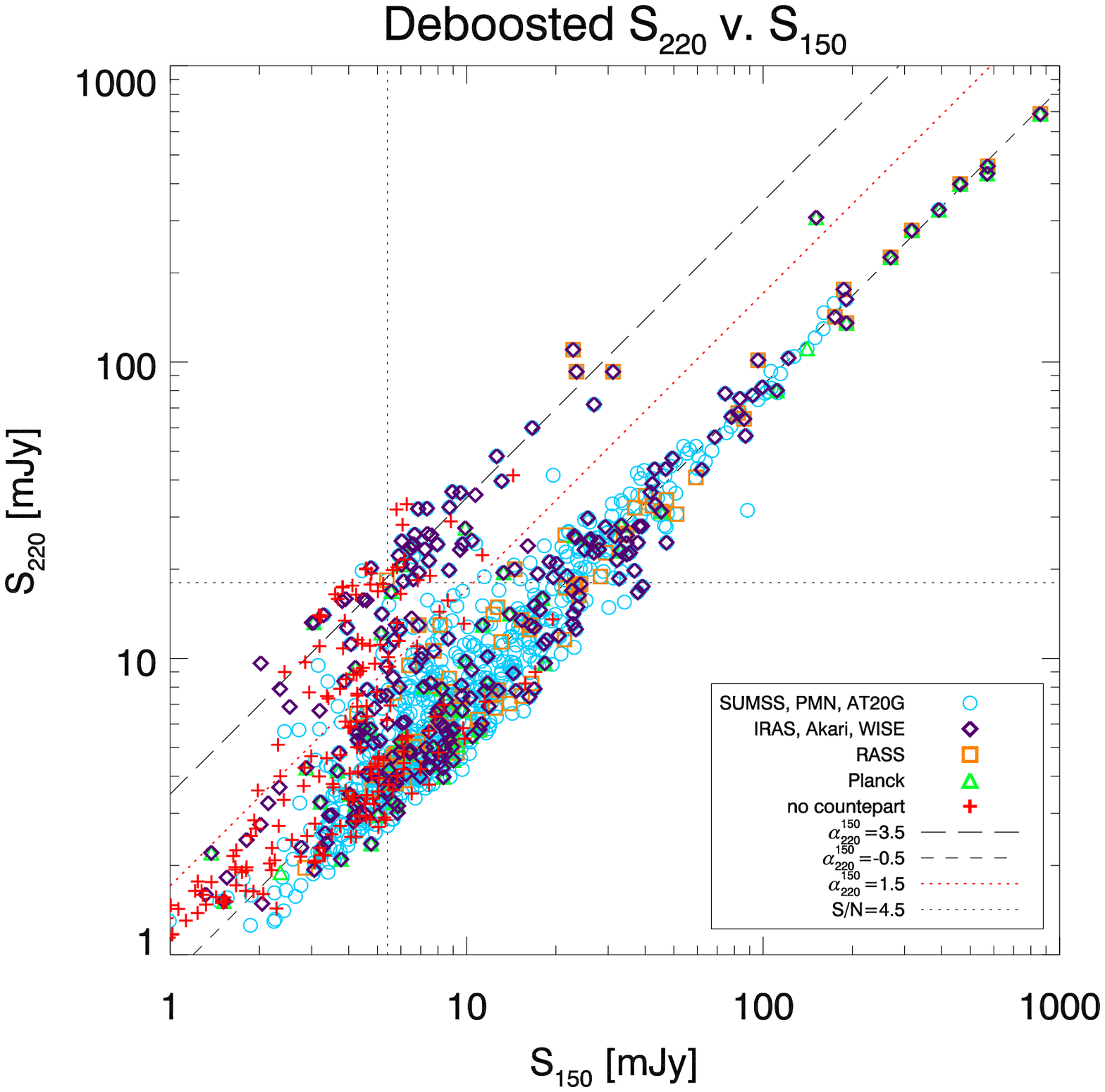}}
\caption{Fluxes in the three bands: $150$~GHz flux versus $95$~GHz flux, raw (top left) and deboosted (top right), and $220$~GHz flux versus $150$~GHz flux, raw (bottom left) and deboosted (bottom right) for sources detected above 4.5$\sigma$ in at least one band. This plot shows the 1128 three-band sources (we leave out the two-band \fiveh~and \twthree~fields). The cyan, purple, orange and green symbols mark SPT-detected sources that have a counterpart in the SUMSS, PMN, AT20G, IRAS, AKARI, WISE, RASS, or Planck catalogs (See \S\ref{sec:extern} for a description of the external catalogs we cross-match against). The red crosses show SPT sources with no matches in these catalogs. The long-dashed line represents the locus for sources with a spectral index $\alpha=3.5$, characteristic of dust emission, while the short-dashed line traces $\alpha=-0.5$, typical of synchrotron-dominated sources. The dotted red line represents the $\alpha=1.5$ threshold for source classification (detailed in \S\ref{sec:alpha}). The dotted black line is the 4.5$\sigma$ detection threshold. \label{fig:rawf}}
\end{center}
\end{figure*}

\subsection{Choice of priors}

Next, we need a prior on $P(\smaxnine,\smaxonef,\smaxtwot)$. Because the fluxes in the three bands are correlated, it is easier to construct a prior for the flux in one band and the two spectral indices of the source, $P(\smaxonef,\alphao,\alphat)$. We employ the simplifying assumption that we can separate this prior as:

\begin{eqnarray}
\label{eq:separate_prior}
P(\smaxonef,\alphao,\alphat) = P(\smaxonef)P(\alphao)P(\alphat).
\end{eqnarray}

For the spectral indices, we use flat priors between $-3$ and 5. The prior $P(\smaxonef)$ is obtained from summing the estimated number counts $dN/dS$ of models of synchrotron and dusty source populations. 
For synchrotron sources, we use the \cite{dezotti05} prediction at 150 GHz, and extrapolate it to the other two bands. 
For dusty sources, we use the Negrello (private communication) predictions at 150 and 220~GHz and extrapolate the \cite{negrello07} predictions at 850~\um\ to the 95 GHz band using a spectral index of 3.1 for SMGs (derived from the Arp 220 SED at a redshift $\sim3$) and 2.0 for the low-redshift ($z<0.3$) IRAS sources. 
We have checked that using different number count models as priors does not significantly impact the final results.

In separating the prior this way, we have assumed above that the spectral index priors do not depend on the source flux and are not correlated. 
In reality, we know that the spectral indices do depend on flux, because the brightest sources are synchrotron-dominated. 
Also, we do expect the two spectral indices to be correlated for most sources, unless there is a strong spectral break between bands. 
However, the $\alpha$ priors are broad enough to let these correlations emerge from the data itself; we choose to stay agnostic about the spectral index distribution and to avoid downweighting potentially different SEDs.

\begin{figure*}[!ht]
\begin{center}
  \subfigure{\includegraphics[width=8.9cm]{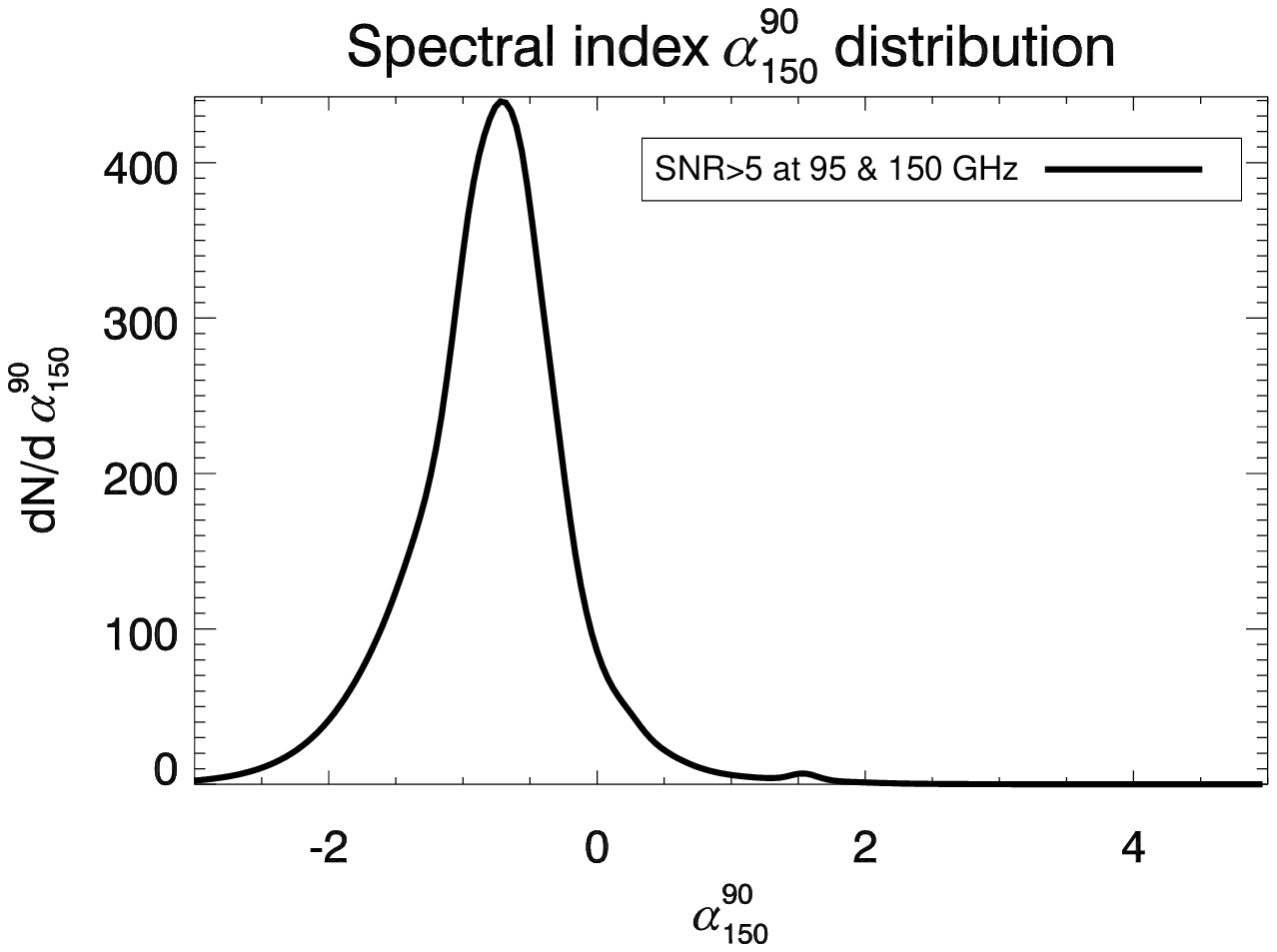}}
  \subfigure{\includegraphics[width=8.9cm]{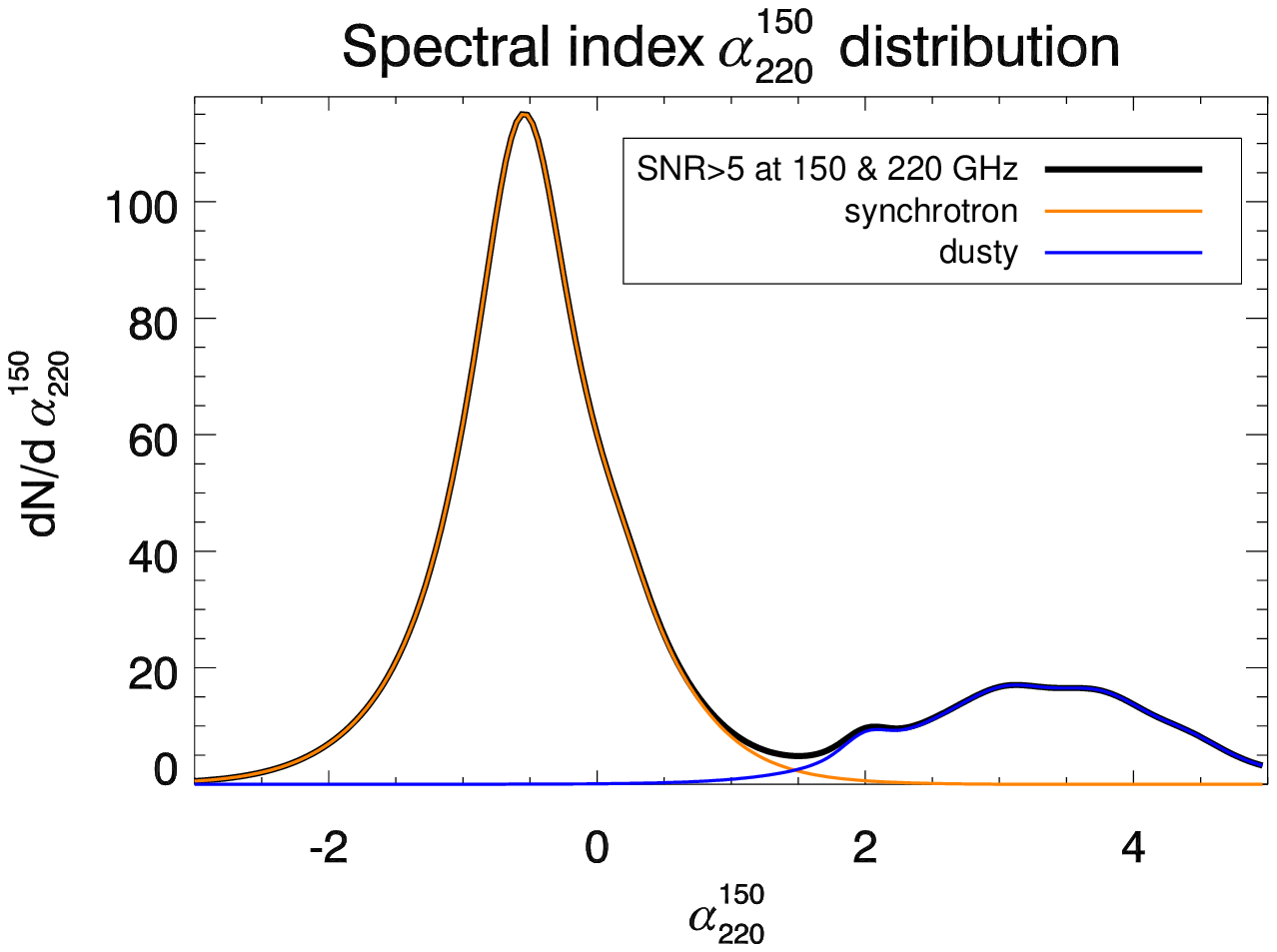}}
\end{center}
\caption{Posterior distribution of spectral indices $\alphao$ (left) and $\alphat$ (right) for sources detected above $5\sigma$ in both adjacent bands that define the respective spectral index. The local minimum of the $\alphat$ distribution, $\alphat$=1.5, is chosen as the threshold for source classification. There are 503 sources, 491 synchrotron and 12 dusty, contributing to the $\alphao$ distribution. There are 191 sources, 151 synchrotron and 40 dusty, contributing to the $\alphat$ distribution. 
\label{fig:a2hist}}
\end{figure*}

The next step is to convert $P(\smaxonef,\alphao,\alphat)$ into a three-flux prior,

\begin{eqnarray}
\label{eq:flux_prior}
P(\smaxnine,\smaxonef,\smaxtwot)  &=&   \nonumber\\ 
P(\smaxonef,\alphao,\alphat)
\left|\frac{d\alphao}{d\smaxnine}\cdot\frac{d\alphat}{d\smaxtwot}\right|.
\end{eqnarray}

We define the ``detection band'' as the band we apply the number counts prior to; prior information in the other two bands is constructed by combining the number counts prior in the detection band with the spectral index priors.  

In the expressions above, 150 GHz is chosen as a detection band. In practice, we perform the deboosting procedure with each band in turn chosen as the detection band by modifying the above calculation accordingly. 
For the fluxes reported in the source catalog, we use the band with the highest significance detection as the detection band. 
When we derive number counts in one band, we use that band as the detection band for all contributing sources.

We note that the three-band deboosting procedure accounts for correlations between bands not only in the prior information, but also in the uncertainty estimates. Beam calibration and the absolute calibration to WMAP are the main sources of band-to-band correlated uncertainty.

\subsection{Posteriors}

We marginalize over the parameters in the three-dimensional posteriors $P(\smaxnine,\smaxonef,\smaxtwot)$ and $P(\smaxonef,\alphao,\alphat)$ to obtain one-dimensional posterior probability densities for the true fluxes $\smaxnine,\smaxonef,\smaxtwot$ and for the spectral indices $\alphao,\alphat$. We take the 16$\%$, 50$\%$, and 84$\%$ values of the cumulative posteriors as the best fit values and equivalent 1$\sigma$ errors.

We construct two distinct sets of posteriors. 
The first set, used for deriving the estimated source fluxes in the catalog, includes all sources of error described above (map noise, beam, and absolute calibration). 
The second set is used for deriving number counts. This set of posteriors does not include the beam and calibration errors, as these two sources of error are common to all the sources in the catalog. We account for those errors by including a common noise realization to all the fluxes in each mock catalog that we construct to obtain statistics; this will be detailed in \S\ref{sec:number_counts}.


\subsection{Deboosted fluxes}
\label{sec:flux}

Figure~\ref{fig:rawf} presents a scatter plot of the fluxes of each source in different bands, for both the raw (left) and deboosted (right) flux values. We note that we only consider sources that have three-band data for this part of the analysis. We thus leave out the \fiveh~and \twthree~fields here. There are several points to note in this figure. 

First, from the bottom panels showing the 220 versus 150 GHz flux, one can note that the sources separate into two populations that roughly follow the loci of spectral index -0.5, typical of synchrotron emission, and 3.5, characteristic of dusty galaxies. The top panels, showing $150$~GHz versus $95$~GHz flux, display many more sources with negative spectral indices, as there are very few sources that are dust-dominated down to 95 GHz.
This figure only gives a rough picture; the actual source classification is based on an integrated posterior probability density function (PDF) of the spectral index and is described in \S \ref{sec:alpha}. Sources that appear below both dotted lines, which are the 4.5$\sigma$ noise threshold levels, are detected only in the band that isn't plotted.

Second, most of the synchrotron sources have counterparts in the SUMSS (or PMN, AT20G) radio catalogs, and roughly half of the dusty sources are in the IRAS (or AKARI, WISE) catalogs (see \S\ref{sec:extern} for a description of the external catalogs that we cross-match against). While most of the sources without counterparts are close to the detection threshold, there exist a number of strongly detected objects of both populations that do not have counterparts. This issue will be explored in \S\ref{sec:extern}.

Third, the figure shows the effect deboosting has on the raw fluxes. The lowest signal-to-noise sources are the most strongly affected, while strong detections show little change.

\begin{figure*}[!ht]
\begin{center}
  \subfigure{\includegraphics[width=8.9cm]{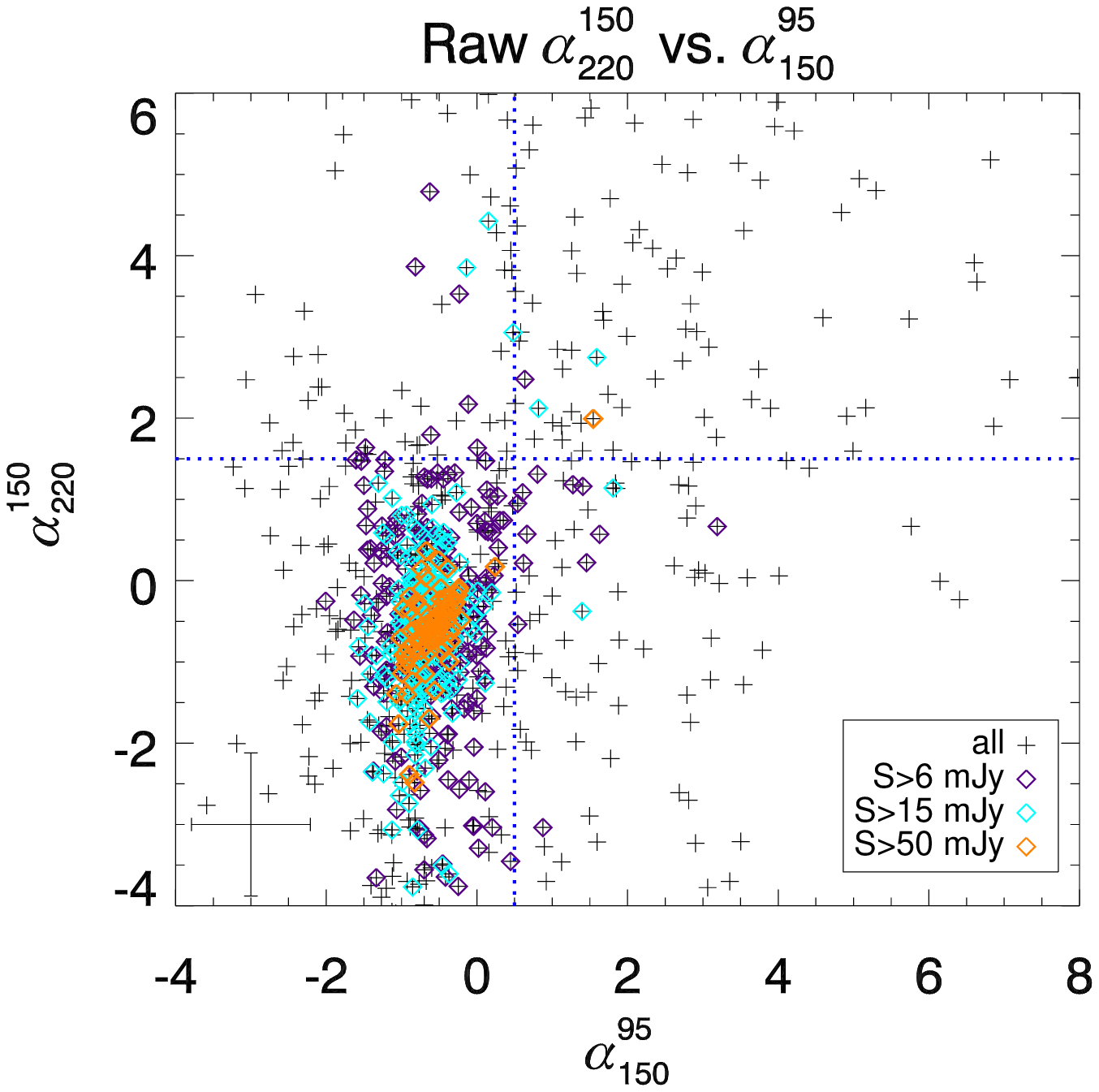}} 
  \subfigure{\includegraphics[width=8.9cm]{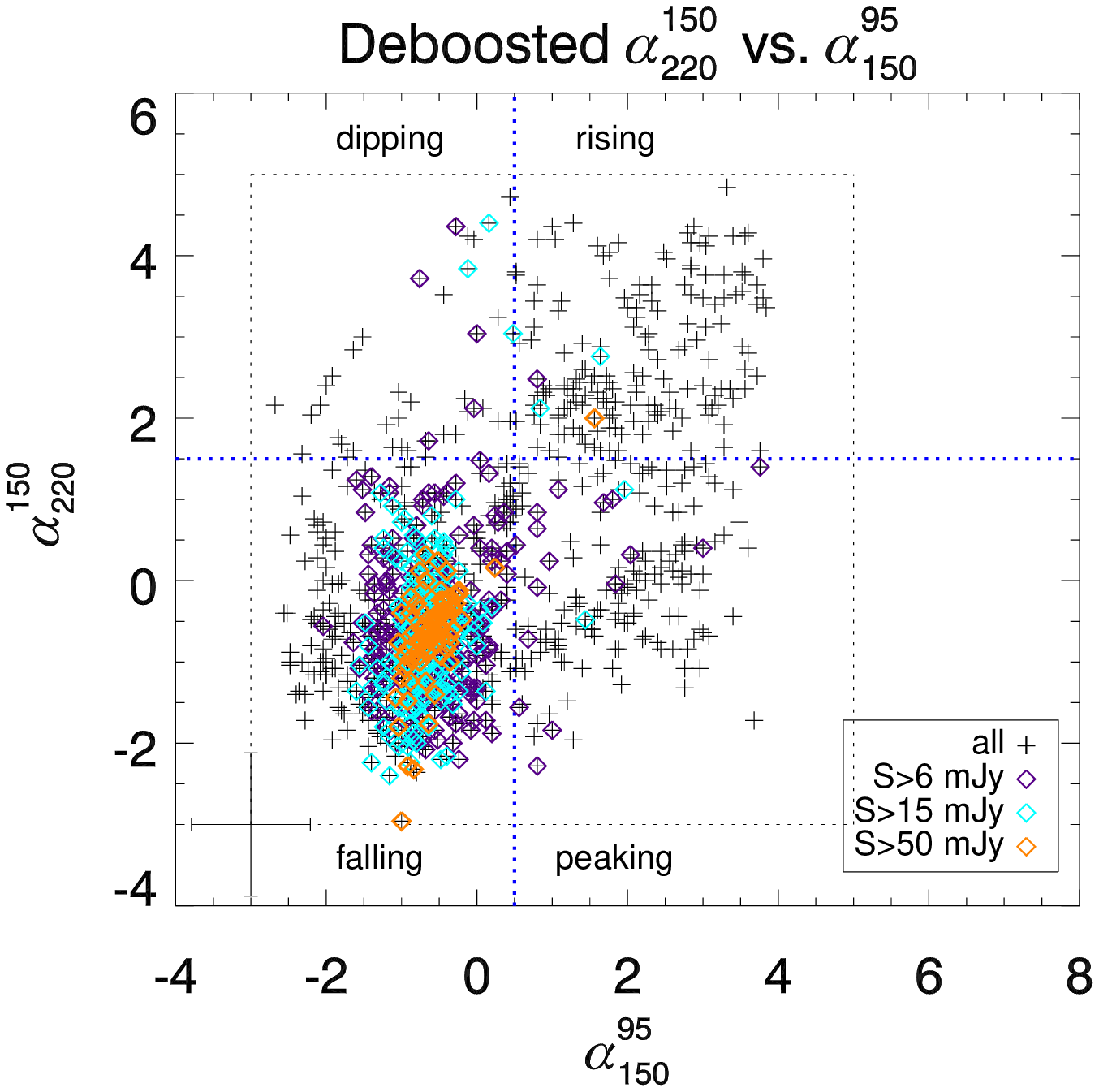}}
  \subfigure{\includegraphics[width=8.9cm]{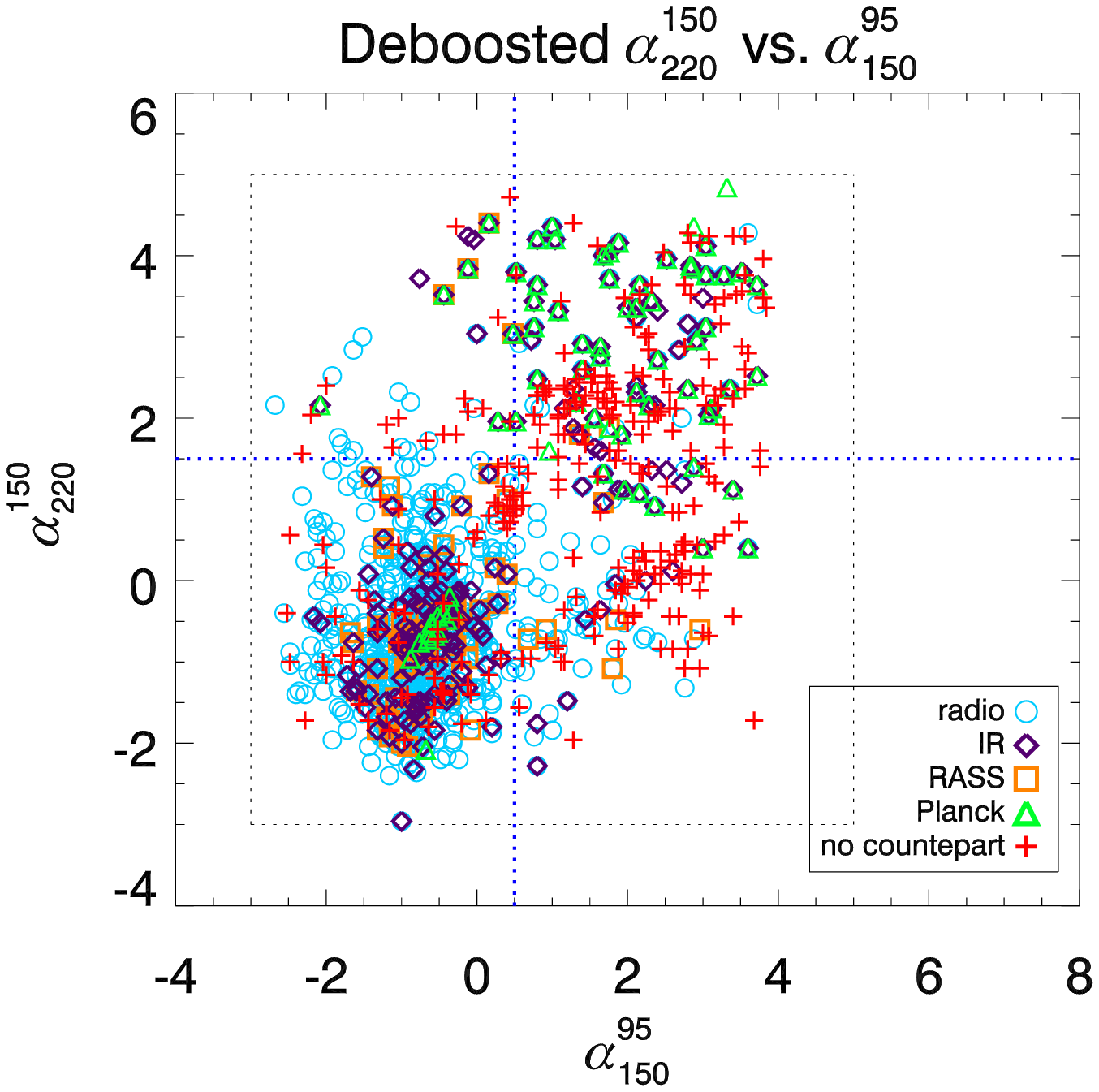}}
\end{center}
\caption{Spectral index $\alphat$ versus $\alphao$ for all 1128 sources in the 2009 fields (we leave out the two-band \fiveh~and \twthree~fields). The top left panel shows the raw spectral indices and the top right panel shows the deboosted values. The color coding refers to sources detected above the flux listed in the legend in at least two bands. The dotted square in the deboosted plot shows the parameter region allowed by the spectral index prior. The dashed blue lines shows the threshold $\alphao$ and $\alphat$ values used as delimiters for the four population quadrants. We note that the actual synchrotron/dusty classification is done probabilistically, based on the posterior PDF, as detailed in \S\ref{sec:alpha}. The crosshair on the bottom left shows typical errors for spectral indices of sources near the detection threshold. The bottom panel shows the deboosted values color-coded based on external catalog counterparts of the sources. ``Radio'' stands for SUMSS, PMN, or AT20G, while ``IR'' stands for IRAS, AKARI, or WISE. 
\label{fig:a1a2}}
\end{figure*}

\subsection{Spectral indices and source classification}
\label{sec:alpha}

We add the spectral index posterior likelihoods for all sources,  normalized such that $\int P(\alpha)d\alpha=1$ for each source, to obtain a distribution of $dN/d\alpha$. Figure~\ref{fig:a2hist} shows the distributions of posterior spectral indices $\alphao$ and $\alphat$ for all sources detected above a S/N of 5 in both adjacent bands that define the spectral index. Again, we only use the three-band data when constructing these plots. The $\alphat$ distribution reveals two source populations, with synchrotron sources peaking around a value of -0.5 and dusty sources around 3.5. 

We choose the local minimum in the $\alphat$ distribution, $\alphat=1.5$, as the threshold for source classification. Sources are classified as synchrotron-dominated if there is a less than 50$\%$ probability that their posterior $\alphat$ is greater than the threshold value, $P(\alphat>1.5)<0.5$, and as dust-dominated if this probability exceeds 50$\%$, $P(\alphat>1.5)\ge 0.5$. We choose this classification criterion because, given the 95 GHz map depth, few dusty sources have a well-measured $\alphao$;
also, the posterior $\alphat$ distribution clearly shows a separation into two populations.

The median spectral index for all 915 three-band synchrotron sources is $\alphatwosync=-0.60$. 
If we restrict the sample to sources detected above 4.5$\sigma$ at both 150 and 220 GHz, $\alphatwosync=-0.52$.
Applying an additional signal-to-noise cut of 5 to the latter criterion, $\alphatwosync=-0.48.$
Thus, the brightest synchrotron-dominated sources appear to have slightly flatter spectra.
The median spectral index for the dusty sources detected above 5$\sigma$ at both 150 and 220 GHz is $\alphatwodust=3.35$.

\renewcommand{\arraystretch}{1.3}
\begin{deluxetable*}  {  l l c c c c c}              %

\tablecaption{Spectral behavior}
\tablehead{
& & & $S_{150}$ &  $S_{150}$ &  $S_{150}$ &  $S_{150}$ \\
Spectral behavior        & & All & $<6$ mJy & 6-12 mJy & 12-36 mJy & $\ge 36$ mJy }
\startdata
Any                                   & &  1128 (100$\%$) & 496 (44.0$\%$) & 335 (29.7\%) & 207 (18.3\%) &  90 (8.0$\%$)\\
&&&&& \\
$\alphao<0.5, \alphat<1.5$  & ``falling''  & 753 (66.8\%) & 206 (41.5$\%$)  & 262 (78.2$\%$) & 196 (94.7$\%$) & 89 (98.9$\%$) \\
$\alphao\ge 0.5, \alphat<1.5$  & ``peaking''  & 162 (14.4\%) & 131 (26.4$\%$) & 29 (8.7$\%$)& 2 (1.0$\%$) & 0 (0.0$\%$) \\
$\alphao<0.5, \alphat\ge 1.5$  & ``dipping''  & 41 (3.6\%)   & 28 (5.6$\%$) & 10 (3.0$\%$) & 3 (1.4$\%$) & 0 (0.0$\%$)\\
$\alphao\ge 0.5, \alphat\ge 1.5$  & ``rising''   & 172 (15.2$\%$) & 131 (26.4$\%$) & 34 (10.1$\%$) & 6 (2.9$\%$) & 1 (1.1$\%$) \\
&&&&& \\
$P(\alphat\ge 1.5)<0.5$ & sync & 915 (81.1$\%$) & 337 (67.9$\%$) & 291 (86.9$\%$) & 298 (95.7$\%$) & 89 (98.9$\%$)\\
$P(\alphat\ge 1.5)>0.5$ & dust & 213 (18.9$\%$) & 159 (32.1$\%$) & 44 (13.1$\%$) & 9 (4.3$\%$) & 1 (1.1$\%$)
\enddata

\label{tab:spectral}
\tablecomments{Distribution of spectral behavior for the 1128 sources that have three-band data.}
\end{deluxetable*}
\renewcommand{\arraystretch}{1}

Figure~\ref{fig:a1a2} shows a scatter plot of the two spectral indices, $\alphat$ versus $\alphao$, for all 1128 sources with three-band data (we leave out the two-band \fiveh~and \twthree~fields). We choose $\alphat=1.5$ (the threshold of synchrotron/dust classification) and $\alphao=0.5$ (by visual inspection of the scatter) as delimiters to split the parameter space into 4 quadrants.  
Table~\ref{tab:spectral} lists the distribution of sources falling into each of those 4 quadrants in several flux bins. 

The two spectral indices show significant correlation, as expected. As listed in Table~\ref{tab:spectral}, the majority of sources have both spectral indices of synchrotron-type (lower left quadrant), consistent with flux that is falling with increasing frequency. This fraction increases from about 40\% for faint sources ($S_{150}<$ 6 mJy) to almost 100\% at the bright end. 
The strongest detected sources are concentrated around the [-0.5, -0.5] point. 

The upper right quadrant of Figure~\ref{fig:a1a2}, encompassing sources with steeply rising flux, is the next most populated. It contains roughly a quarter of the fainter sources ($S_{150}<$ 6 mJy), but the fraction drops significantly at bright fluxes.

We also detect sources with high $\alphao$ and synchrotron-type $\alphat$ (lower right quadrant)---suggesting a peaking or flattening spectrum between 95 and 150 GHz---and sources with low $\alphao$ and dust-like $\alphat$ (upper left quadrant), consistent with spectra that have a minimum between 95 and 220 GHz. These populations will be discussed in more detail in \S\ref{sec:pop}.

\section{Catalog description, statistics and external associations}
\label{sec:catalog}

We detect 1545 sources above 4.5$\sigma$ and 1109 above 5$\sigma$ in any one band. Of these, 
1238 above 4.5$\sigma$ (964 above 5$\sigma$), or 80.1$\%$ (86.9$\%$) are classified as synchrotron-dominated,
and 307 (145), or 19.9$\%$ (13.1$\%$) are classified as dust-dominated. 

The map pixel flux histograms are well approximated by Gaussian distributions. The map RMS is roughly $2.1$~mJy at $95$~GHz, $1.2$~mJy at $150$~GHz, and $3.9$~mJy at $220$~GHz (see Table~\ref{tab:completeness}), with a mild dependence on declination within a given field (the noise is slightly lower at more negative declinations). We note that the field depths vary slightly as a function of observing time and focal plane configuration for each year.

In the 2009 fields, totalling 584 \sqdeg~and having three-band data, we detect 640 sources in the 95 GHz map, 915 at 150 GHz, and 344 at 220 GHz.
In the two 2008 fields with two-band data, we detect 331 sources at 150 GHz and 191 at 220 GHz. 
After combining the single-band catalogs, we are left with a total of 1545 (1109) sources detected above 4.5$\sigma$ (5$\sigma$) in at least one band. Of those, 1128 (816) are from the 2009 data and have three-band information, and 417 (293) are from the 2008 fields and only have two-band information.

\subsection{Catalog description}

We construct a catalog with the following entries:
\begin{enumerate}
\item Source ID:  the IAU designation for the SPT-detected source.
\item RA:  right ascension (J2000) in degrees.  
\item DEC:  declination (J2000) in degrees.
\item $S/N_{95}$:  detection significance (signal-to-noise ratio) in the $95$~GHz band.
\item $S_{95}^\mathrm{raw}$: raw flux (uncorrected for flux boosting) in the $95$~GHz band.
\item $S_{95}^\mathrm{dist}$: deboosted flux values encompassing $16\%$, $50\%$, and $84\%$ ($68\%$ probability enclosed, or $1\sigma$ for the equivalent normal distribution) of the cumulative posterior probability density for $95$~GHz flux, as estimated using the deboosting procedure described in \S\ref{sec:deboost}.
\item $S/N_{150}$: detection significance at $150$~GHz.
\item $S_{150}^\mathrm{raw}$: raw flux at $150$~GHz.
\item $S_{150}^\mathrm{dist}$: deboosted flux values at $150$~GHz.
\item $S/N_{220}$: detection significance at $220$~GHz.
\item $S_{220}^\mathrm{raw}$: raw flux at $220$~GHz.
\item $S_{220}^\mathrm{dist}$: deboosted flux values at $220$~GHz.
\item $\alphaor$: estimate (from the raw flux in each band) of the $95$~GHz$-$$150$~GHz spectral index $\alphao$.
\item $\alphaod$:  $16\%$, $50\%$, and $84\%$ estimates of the spectral index, based on the posterior probability densities for the spectral index calculated using the deboosting procedure described in \S\ref{sec:deboost}.
\item $\alphatr$: estimate (from the raw flux in each band) of the $150$~GHz$-$$220$~GHz spectral index $\alphat$.
\item $\alphatd$:  $16\%$, $50\%$, and $84\%$ estimates of the spectral index calculated using the deboosting procedure.
\item $P(\alphat > \alphathresh)$: fraction of the spectral index posterior probability density above the threshold value of $\alphathresh$. A higher value of $P$ means the source is more likely to be dust-dominated. This is detailed in \ref{sec:alpha}.
\item Type: source classification (synchrotron- or dust-dominated), 
based on whether $P(\alphat > \alphathresh)$ is greater than or less than $0.5$. 
\item External counterparts: external catalogs wherein a source has a match with an offset smaller than the chosen association radius. As described in \S\ref{sec:extern}, we choose an association radius of 1 arcminute for all catalogs except WISE, where we use 0.5 arcminutes.
\item Extended flag: flag for extended sources.

\end{enumerate}

The catalog is available for download on the SPT website.\footnote{http://pole.uchicago.edu/public/data/mocanu13/}

\subsection{Completeness}
\label{sec:completeness}

To estimate the completeness of the catalog, we check how well the source-finding algorithm detects a known sample of sources. For this purpose, we take the residual map for each field, which is a good approximation of noise, and add simulated sources of a fixed flux at random locations. We construct the simulated source profiles from the measured beam convolved with the map-domain equivalent of timestream filtering and matched filter. This is equivalent to the source profile described in \S\ref{sec:clean}. We then run the source-finder on those maps to find the number of input sources that are recovered as a function of flux. It follows that the completeness is $f_\mathrm{compl}(S) = N_\mathrm{recovered}/N_\mathrm{input}$. As noise in the maps is Gaussian and uniform to a good approximation, the cumulative completeness is well fit by an error function
\begin{equation}
f_\mathrm{compl}(S) = \frac{1}{\sqrt{2 \pi \sigma^2}} \int_{S}^\infty e^{-\left(S^\prime - S_0 \right)^2/2\sigma^2}dS^\prime,
\label{eqn:compl}
\end{equation}
where $S_0$ is the detection threshold. We find the best-fit $\sigma$ value for each field and band and use this function as an estimate of completeness. The $50\%$ completeness levels are, on average, 9.1 mJy, 5.4 mJy, and 17.6 mJy at 95, 150, and 220 GHz respectively. We are $95\%$ complete at roughly 12.6, 7.4, and 24.1 mJy at 95, 150, and 220 GHz respectively. 
Table~\ref{tab:completeness} shows the depth and the 50\% and 95\% completeness levels for each field.

\begin{deluxetable*}{ l | c c c |c c c| c c c}
\tablecaption{Field depths and completeness levels} \small

\tablehead{
& & 95 GHz & & & 150 GHz & & & 220 GHz &\\
Name & RMS & 50\% c. & 95\% c. & RMS & 50\% c. & 95\% c. & RMS & 50\% c. & 95\% c. \\
& (mJy) & (mJy) & (mJy) & (mJy) & (mJy)  & (mJy) & (mJy) & (mJy)  & (mJy) }
\startdata
 
\fiveh        &      & - &            & 1.27  & 5.25 & 8.25 & 3.35 & 13.65 & 21.15 \\
\twthree      &      & - &            & 1.24  & 5.40 & 7.38 &  3.56 & 15.75 & 21.51 \\
\twonesix     & 1.95  & 8.55 & 11.68  & 1.13  &  4.95 & 6.76 & 3.94 &  17.55 & 23.97 \\
\three        & 2.04  & 8.91 & 12.17  & 1.19 & 5.54 & 7.57 & 4.02 & 17.86 & 24.40 \\
\twonefif     & 2.27 & 9.86 & 13.46   & 1.32 & 5.85 & 7.99 & 4.49 &  19.62 & 26.79

\enddata

\label{tab:completeness}
\tablecomments{RMS noise and 50\% and 95\% completeness levels for each field.}
\end{deluxetable*}

We note that 
galaxy clusters with very significant and compact SZ decrements  
are detected and CLEANed in the source-finding process. 
We are thus not accounting for any incompleteness caused by
sources having their emission cancelled by the decrements from these clusters. 
However, assuming a WMAP7 cosmology, a \citet{tinker08} cluster mass 
function, and the SPT cluster mass and redshift selection function from \citet{reichardt13}, 
we expect roughly one decrement large enough to cancel a $4.5 \sigma$ point 
source per ten square degrees, or roughly 80 in the entire area used here.
If there is no spatial correlation between clusters and point sources, then the probability that even 
one $>$4.5$\sigma$ point source is being cancelled by one of these clusters is very small, roughly 1\%.
There is, of course, theoretical motivation, as well as some observational evidence, for a correlation between
clusters and point sources.  
But even if every cluster we remove is hiding a 
$4.5\sigma$ source---which is an extremely pessimistic upper limit---this 
will cause only a few-percent error in the completeness calculation at 95 and 
150~GHz. (The 220~GHz completeness calculation is unaffected by SZ.)

\subsection{Purity}
\label{sec:purity}

We estimate purity by running the source-finder on simulated maps. 
These maps are constructed by taking difference maps that contain only atmospheric and instrumental noise and adding a CMB realization from the best fit WMAP7 + K11 CMB power spectrum, estimates of the Sunyaev-Zel'dovich (SZ) effect, and Poisson and correlated components of the CIB. 
We calculate the purity fraction as a function of signal-to-noise as $f_\mathrm{pure}=1- N_\mathrm{false} / N_\mathrm{total}$, where $N_\mathrm{false}$ is the number of detections in the simulated maps above a certain signal-to-noise, and $N_\mathrm{total}$ is the total number of sources detected in the real maps above the same threshold. We find the catalog to be 92\% pure at 4.5$\sigma$ in the 150 GHz band.

The simulations used to calculate purity do not include the SZ effect from massive clusters; the SZ 
effect in the simulations is a Gaussian field with the power spectrum tuned to match the measurement
in \citet{shirokoff11}. Thus we are not accounting in the purity calculation for possible spurious positive
source detections from the wings of very significant (negative) SZ decrements. In practice, such spurious
detections are both very rare and easily detectable, so we can remove them from the catalog if necessary.
The most significant cluster in these fields is SPT-CL~J2106-5844, which is also among the most compact
due to its very high redshift ($z=1.18$, \citealt{foley11}), making it the most likely source of detectable
positive wings. This cluster's decrement produces a $5.0\sigma$ wing at 150~GHz and a $5.4\sigma$
wing at 95~GHz, and we remove these spurious detections from the catalog. 
The next most significant cluster in these fields is a factor of 1.5 less significant than
SPT-CL~J2106-5844 \citep{reichardt13}, so we do not expect any other clusters to produce detectable
positive wings from their SZ decrements.

\subsection{Extended sources}

Given the SPT's arcminute resolution, extragalactic sources at redshifts above $z \sim 0.05$ are expected to appear pointlike in the maps. Only very nearby sources or AGN with extended structure (radio lobes or jets) are expected to look extended.

We test all sources detected above a signal-to-noise of 5 in any band for extended emission. We take a cut-out of an unfiltered map around each detected source and fit to it the measured beam convolved with a two-dimensional elliptical Gaussian function, letting the width along two directions and the orientation angle vary. Based on an empirical comparison of \delchisq~for the extended model and visual evidence for extended emission, we chose to flag as extended those sources for which $\delchisq>10$ between the best fit extended model and the beam-only model.

The flux calculation uses a source template which consists of a filtered beam. The flux of an extended object will be underestimated because the effective solid angle under the source template that we use in the flux calculation (Equation \ref{eqn:area_eff}) corresponds to a point source.

We detect 63 extended sources, out of which 37 are synchrotron-dominated and 26 are dust-dominated. The brightest extended sources are AGN with extended emission, generally due to lobe structures. This is confirmed by their extended, multiple-blob or jet-like appearance in the corresponding SUMSS image, which we visually inspect for the brightest 20 sources. The brightest dusty extended sources are nearby star-forming galaxies present in the New General Catalogue of Nebulae and Clusters of Stars (NGC). Some examples are NGC 1599, NGC 1672 (Seyfert type 2 nucleus, with strong and extended emission in both radio and infrared), NGC 1566 (the second brightest known Seyfert galaxy, which also appears extended in SUMSS maps), and NGC 7090. Fainter detections include NGC 7083, 7059, 7125 and 7126. All of the extended sources we find have counterparts in external catalogs.

\subsection{External associations}
\label{sec:extern}

\begin{deluxetable*}{l|c c c c c}[!h]
\tablecolumns{6}
\tablecaption{Counterparts in external catalogs}
\tablehead{ \colhead{Name} & \colhead{No. counterparts} & \colhead{No. footpr.} & \colhead{$\Sigma$ (per \sqdeg)} & \colhead{$r_{assoc}$ (arcmin)} & \colhead{P(random) ($\%$)} }
\startdata
SUMSS    & 1092 & 22538 & 29.23 & 1 & 2.55 \\
IRAS     & 98 & 4349 & 5.64 & 1 & 0.49 \\
RASS     & 113 & 2718 & 3.53 & 1 & 0.31 \\
AKARI    & 82 & 5583 & 7.24 & 1 & 0.63 \\
Planck   & 101 & 587 & 0.76 & 1 & 0.07 \\
WISE     & 274 & 54005 &  70.05 & 0.5 & 1.52 \\
PMN      & 530 & 1562 & 2.03 & 1 & 0.18 \\
AT20G    & 277 & 297 & 0.39 & 1 & 0.03

\enddata

\label{tab:external}
\tablecomments{Summary of cross-matching with external catalogs. The table lists the catalog name, number of SPT sources with counterparts in that catalog within the listed association radius, the number of sources in that catalog located within the 5 SPT fields, source density within the SPT fields, chosen association radius, and the probability of random association with an SPT source given the association radius.}
\end{deluxetable*}

We search several external catalogs for counterparts at the positions of all sources in the catalog. We query the following catalogs:

\begin{enumerate}
\item The Sydney University Molonglo Sky Survey \citep[SUMSS,][]{mauch03} at 36 cm (843 MHz).
\item The Parkes-MIT-NRAO \citep[PMN,][]{wright94} Southern Survey at 4850 MHz.
\item The Australia Telescope 20 GHz Survey \citep[AT20G,][]{murphy10} at 1.5 cm.
\item The IRAS Faint Source Catalog \citep[IRAS-FSC,][]{moshir92} at 60 and 100~\um.
\item The WISE Source Catalog at 22~\um~(W4).
\item The AKARI/FIS Bright Source Catalog \citep{yamamura10} at 65, 90, 140 and 160 \um~and AKARI/IRC Point Source Catalog \citep{ishihara10} at 9 and 18~\um.
\item The Planck Catalog of Compact Sources \citep[PCCS,][]{planck13-28} at 30, 44, 70, 100, 143, 217, 353, 545, 857 GHz (1 cm to 350 \um).

\item The ROSAT All-Sky Survey (RASS) Bright Source Catalog \citep{voges99} and Faint Source Catalog \citep{voges00}.
\end{enumerate}

These are the most relevant catalogs to search for millimeter-wave selected extragalactic sources in the Southern Hemisphere. SUMSS is the essential radio catalog to check, as it has complete coverage of the SPT fields to a $5\sigma$ depth of 6 mJy/beam. For this reason, we expect most of our significant synchrotron-dominated sources to have counterparts in SUMSS. We add PMN and AT20G in the radio catalog category. The IRAS and AKARI catalogs are the longest-wavelength infrared catalogs with full-sky coverage and thus the most appropriate catalogs to check for local dust emission from LIRGs and ULIRGs. We also check the WISE and Planck catalogs for potential dusty-source counterparts. 

We use an association radius of one arcminute for all catalogs except WISE, for which we use 30 arcseconds. These values were chosen based on the positional accuracy of the catalogs and their beam size and by looking at the distribution of offsets between SPT sources and their closest counterpart in each catalog as a function of SPT signal-to-noise. 

Table~\ref{tab:external} shows the number of SPT sources with counterparts in each catalog, the source density and probability of random association for the chosen radius for each catalog. We note that nearly all of the dusty sources associated with WISE or AKARI are also in IRAS, and nearly all sources found in PMN or AT20G are also found in SUMSS. 

\begin{deluxetable*}{  l | c c c c  }
\tablecaption{Sources without counterparts}
\tablehead{
& &  SNR & (any band) &\\
\colhead{Category} & \colhead{ $>$4.5} & \colhead{$>$5} & \colhead{$>$7} & \colhead{$>$10}  }
\startdata

Any             & 378/ 244/ 1545 & 129/ 28/ 1109 & 20/ 0/ 638 & 6/ 0/ 433  \\
Synchrotron     & 189/ 195/ 1238 & 68/ 24/ 964  & 13/ 0/ 599 & 4/ 0/ 419   \\
Dust            & 189/ 49/ 307 & 61/ 4/ 145  & 7/ 0/ 39  & 2/ 0/ 14   \\
SMG             & 137/ 36/ 174 & 57/ 4/ 80   & 5/ 0/ 13  & 2/ 0/ 5 

\enddata
\tablecomments{Number of sources without counterparts/ expected number of false detections/ total sources in each specified category above the listed signal-to-noise level in any one band.}

\label{tab:counterp}
\end{deluxetable*}


\begin{figure*}
\begin{center}
  \subfigure{\includegraphics[width=8.9cm]{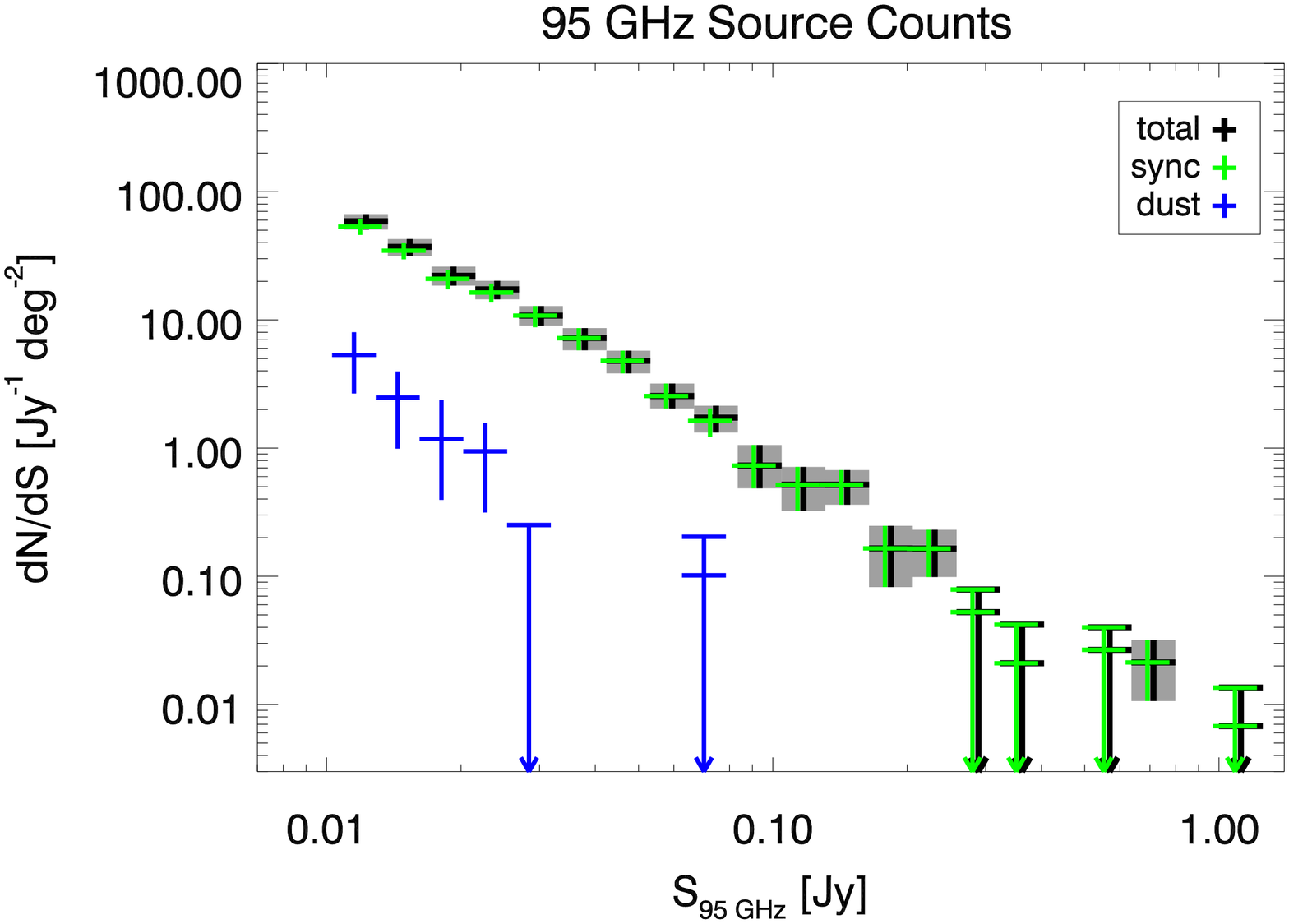}}
  \subfigure{\includegraphics[width=8.9cm]{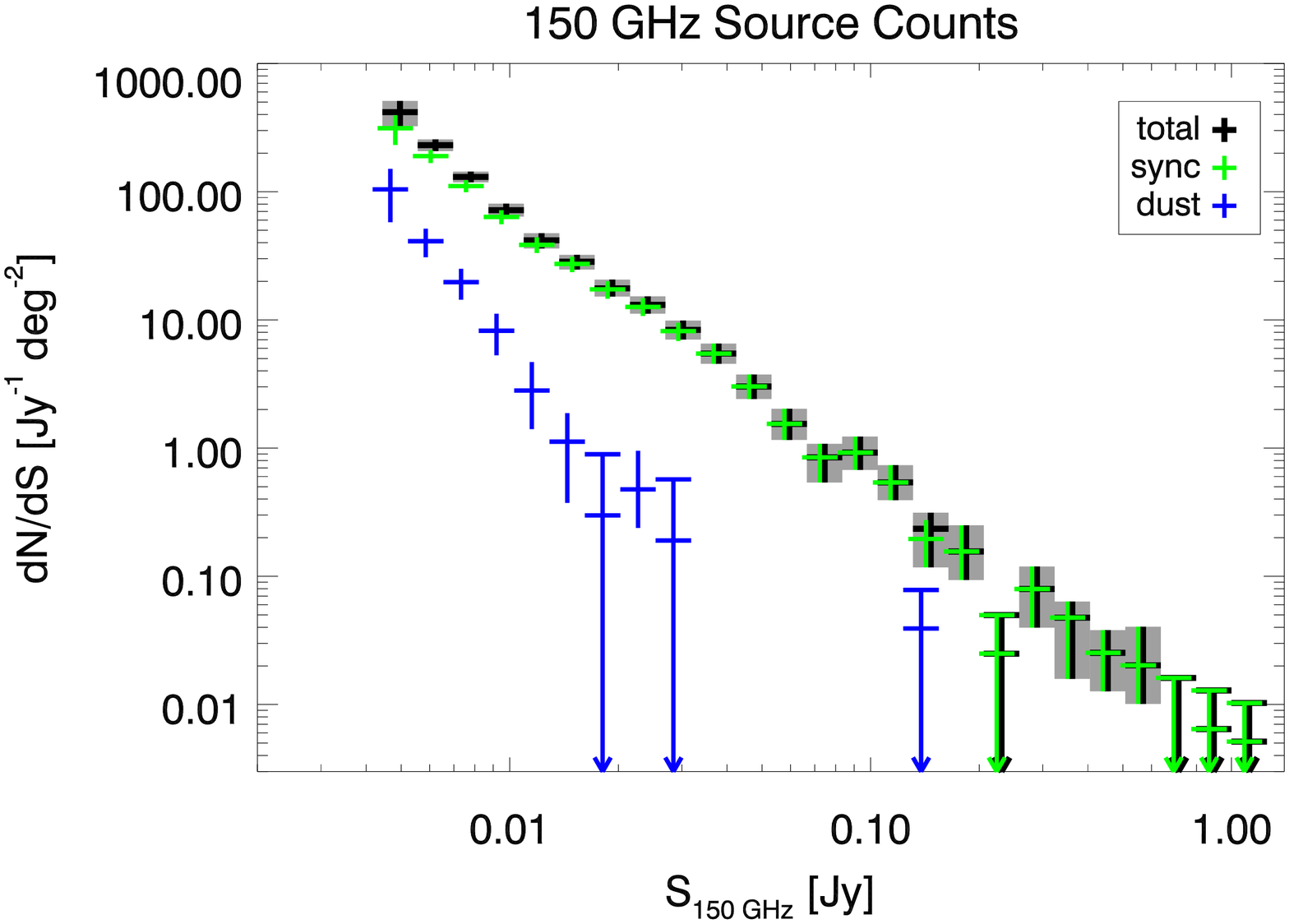}}
  \subfigure{\includegraphics[width=8.9cm]{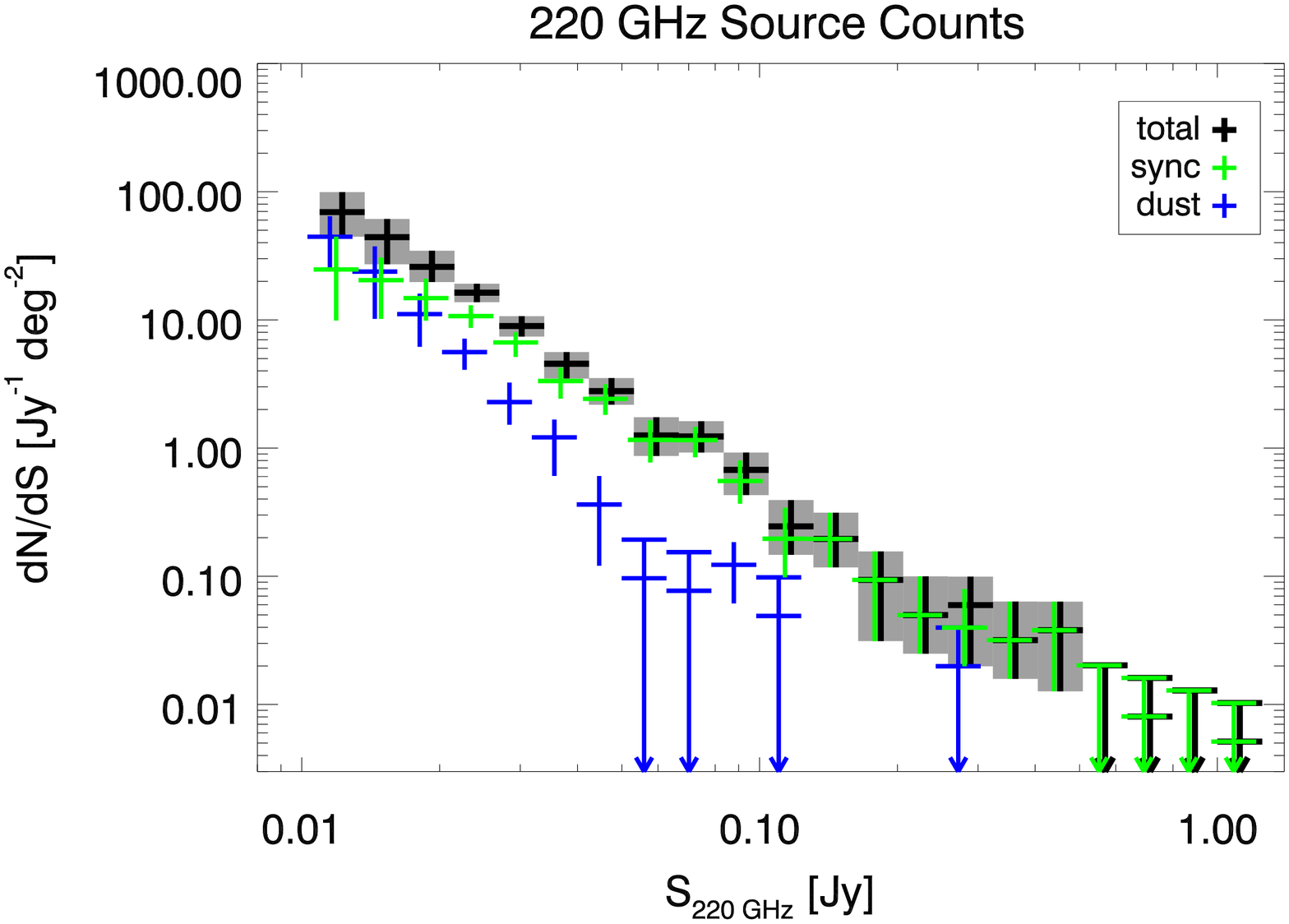}}
\end{center}
\caption{Differential number counts of emissive sources in the three SPT survey bands. Total counts are shown in black, synchrotron-dominated counts are in green, and dust-dominated counts are in blue.
\label{fig:counts}}
\end{figure*}

\begin{figure*}
\begin{center}
  \subfigure{\includegraphics[width=8.9cm]{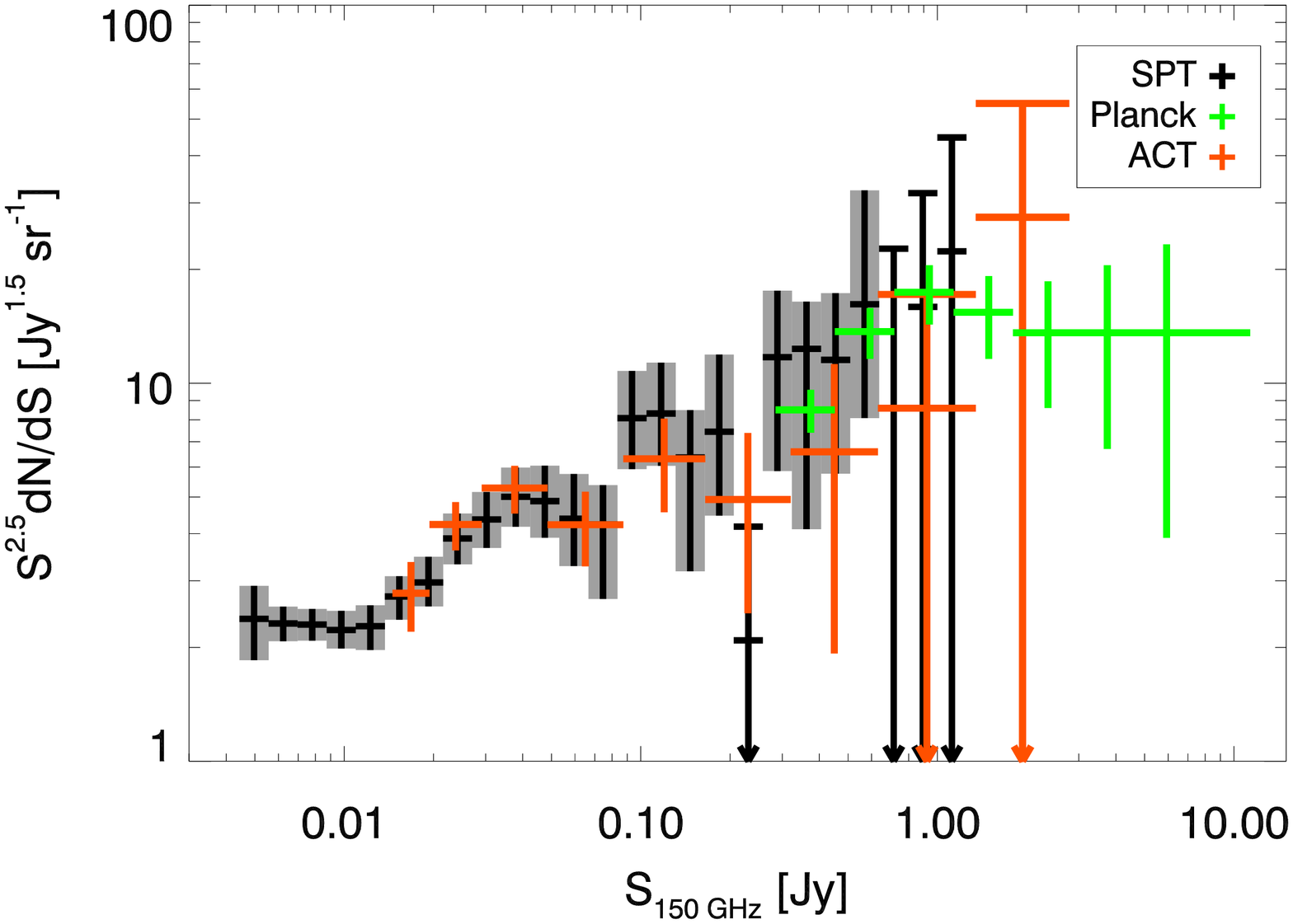}}
  \subfigure{\includegraphics[width=8.9cm]{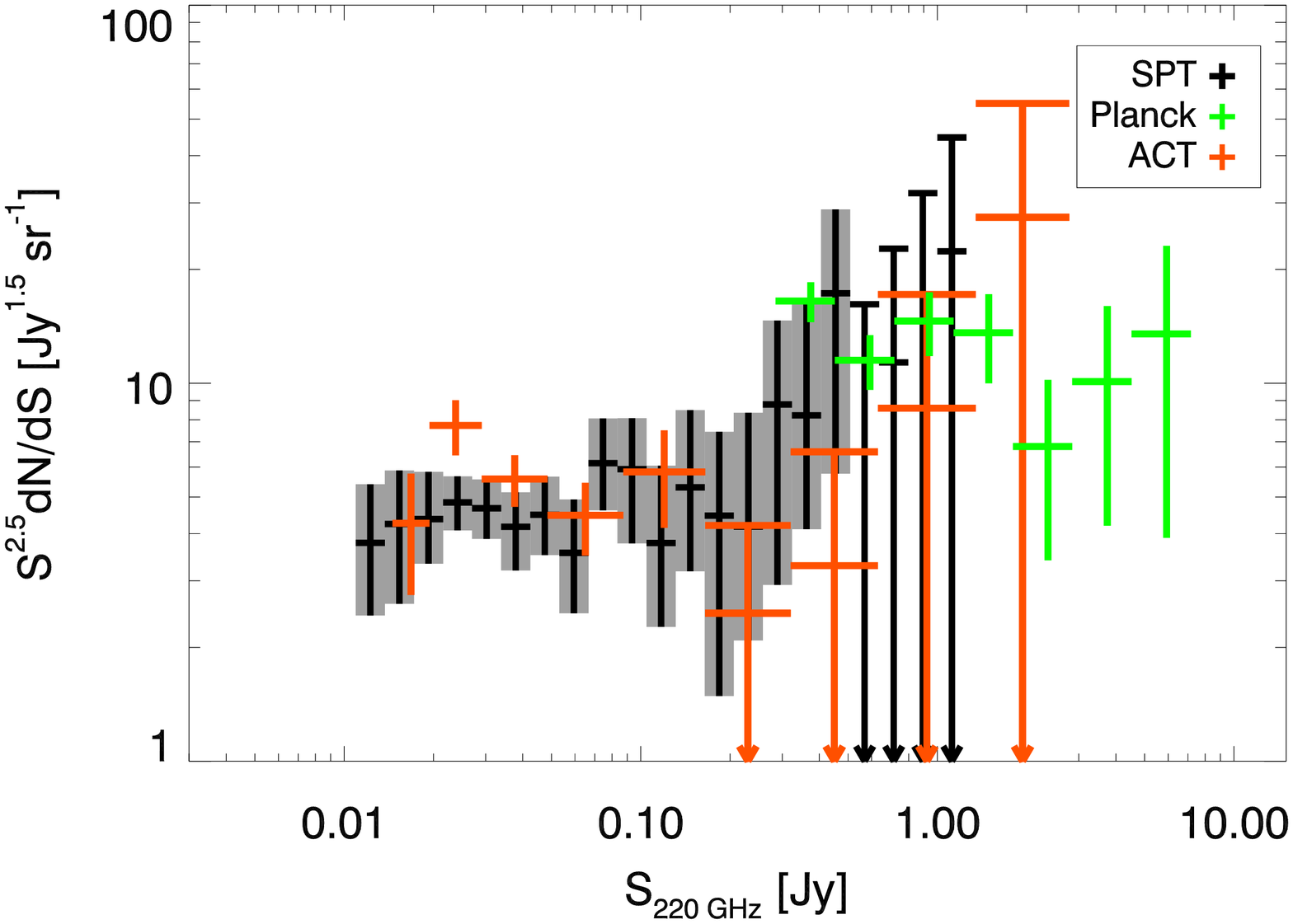}}
\end{center}
\caption{Number counts at 150 and 220 GHz from SPT (this work), Planck \citep{planck12-7}, and ACT \citep{marsden13}.}
\label{fig:comparison}
\end{figure*}

Previously unidentified sources are of particular interest. Table~\ref{tab:counterp} lists the number of SPT sources with no counterparts in any of the catalogs listed above, in total and by category, as well as the expected number of false detections (from the simulations used in \S\ref{sec:purity}), given the signficance level and total number of detections. The synchrotron/dusty classification is done as described in \S\ref{sec:alpha}. We add a subcategory of dusty sources, labelled as ``SMGs'', which we define to be the sources with $\alphat\ge 2$ that have no IRAS counterparts. This is the subset of sources which \citet{vieira13} have demonstrated to have a high probability of being high-redshift, strongly lensed galaxies. We note, however, that the IRAS sky coverage is not perfect and there is a possibility that a few low-redshift objects have been missed by the survey.

We find that almost 25\% (12\%) of all sources above 4.5$\sigma$ (5$\sigma$) do not have counterparts, with 15\% (7\%) of radio sources, 62\% (42\%) of dusty sources and 79\% (71\%) of SMGs lacking external associations. 
As shown in Table~\ref{tab:counterp}, a substantial fraction of sources below $5\sigma$ without counterparts---particularly the synchrotron-dominated sources---are likely to be false detections; above $5\sigma$, however, most sources without counterparts in all categories are expected to be real.
There are 56 synchrotron sources detected above 5$\sigma$ at 150 GHz that have no counterparts in external catalogs. This number is rather surprising, given that basically all synchrotron sources above 5$\sigma$ published in V10 had external associations.

Is it plausible that the SPT could detect synchrotron-dominated sources that were not detected in past radio surveys?
Radio sources are known for their variability, so the synchrotron sources without counterparts in radio catalogs might be flaring between the two observation epochs. The SUMSS detection threshold is between 6 and 10 mJy and the catalog is complete at 18 mJy. A source detected at 5$\sigma$ at 95 GHz in the SPT survey has a flux of about 10 mJy. Assuming a spectral index of -0.5, this means that its SUMSS flux should be around 100 mJy. Therefore, the source could have escaped a SUMSS detection if it flared by a factor of 10 at the time of the SPT detection, which is a reasonable factor (see, e.g., \citealt{aller11}). 

Alternatively, given that some of these sources are only faintly detected at 95 or 220 GHz and their spectral index posteriors are quite wide, it could also be the case that some fraction of them have been misclassified as radio sources. Another possibility is that some faint sources in the SUMSS catalog may have been removed in error by the decision tree used for source selection (according to the SUMSS documentation, \citealt{mauch03}); yet, this is unlikely to have affected more than a few sources.

For any of these explanations, the 7.9 times larger area used in this work makes it more likely to find anomalous sources compared to the V10 analysis. 
However, even accounting for the area differences, the results show some tension with V10. Considering just the SUMSS catalog, there are 73 synchtrotron-dominated sources detected  above 5$\sigma$ at 150 GHz without a counterpart in the four fields analyzed here, and their number in each field is roughly proportional to the area of the field. 
In retrospect, using just the 2009 results, we would predict 11.3 synchrotron-dominated sources without a SUMSS counterpart above 5$\sigma$ in the \twthree~field, and 9.8 such sources in the V10 (\fiveh) field. In reality, we see 7 such sources in the \twthree~field, which is consistent with the prediction, but there is only one such source in the V10 field. Under the assumption of pure Poisson statistics, we would expect one such source or fewer in a field the size of the V10 field less than $0.5 \%$ of the time.

We can ask whether there is something particular about the V10 field that would make it less likely to harbor synchrotron-dominated sources with no SUMSS counterparts.
The \twthree~field is effectively the same as the V10 field in terms of number of bands and depth, so any difference in the V10 field is not due to having three-band data or less deep 220 GHz data. 
We have checked that the SUMSS source densities are similar in the 5 fields. We conclude that the discrepancy between the V10 field and the other four fields is likely a statistical fluctuation.

The dusty sources without counterparts are likely high-redshift galaxies, given that nearby objects would be detected by IRAS. These are interesting sources to follow up and constitute good candidates for strongly lensed SMGs. Some of the brightest such detections in the survey have already been followed up, as noted in \S\ref{sec:intro}, and have been found to indeed be strongly lensed.


\begin{deluxetable*}{llllc}
\tablecaption{The $95$~GHz differential counts.
\label{tab:diffcountstable90}}
\tablehead{
\colhead{Flux range} & \colhead{$dN/dS$ total} & \colhead{$dN/dS$ sync} & \colhead{$dN/dS$ dust} & \colhead{completeness} \\
\colhead{Jy} & \colhead{${\rm Jy}^{-1} {\rm deg}^{-2}$} & \colhead{${\rm Jy}^{-1} {\rm deg}^{-2}$} & \colhead{${\rm Jy}^{-1} {\rm deg}^{-2}$} & 
}
\startdata
$1.1\times10^{-2}-1.4\times10^{-2}$ & $(5.87_{-0.8}^{+0.8})\times 10^{1}$ & $(5.33_{-0.7}^{+0.8})\times 10^{1}$ & $5.33_{-2.7}^{+2.7}$ & $0.93$\\
$1.4\times10^{-2}-1.7\times10^{-2}$ & $(3.71_{-0.5}^{+0.5})\times 10^{1}$ & $(3.47_{-0.5}^{+0.5})\times 10^{1}$ & $2.48_{-1.5}^{+1.5}$ & $1.00$\\
$1.7\times10^{-2}-2.2\times10^{-2}$ & $(2.21_{-0.4}^{+0.4})\times 10^{1}$ & $(2.09_{-0.4}^{+0.4})\times 10^{1}$ & $1.18_{-0.8}^{+1.2}$ & $1.00$\\
$2.2\times10^{-2}-2.7\times10^{-2}$ & $(1.73_{-0.3}^{+0.3})\times 10^{1}$ & $(1.64_{-0.3}^{+0.3})\times 10^{1}$ & $(9.44_{-6.3}^{+6.3})\times 10^{-1}$ & $1.00$\\
$2.7\times10^{-2}-3.4\times10^{-2}$ & $(1.08_{-0.2}^{+0.2})\times 10^{1}$ & $(1.08_{-0.2}^{+0.2})\times 10^{1}$ & $ 0 _{- 0 }^{+0.3}$ & $1.00$\\
$3.4\times10^{-2}-4.2\times10^{-2}$ & $7.21_{-1.4}^{+1.4}$ & $7.21_{-1.4}^{+1.4}$ &   & $1.00$\\
$4.2\times10^{-2}-5.3\times10^{-2}$ & $4.79_{-1.0}^{+1.0}$ & $4.79_{-1.0}^{+1.0}$ &   & $1.00$\\
$5.3\times10^{-2}-6.7\times10^{-2}$ & $2.55_{-0.5}^{+0.6}$ & $2.55_{-0.5}^{+0.6}$ &   & $1.00$\\
$6.7\times10^{-2}-8.3\times10^{-2}$ & $1.73_{-0.4}^{+0.4}$ & $1.63_{-0.4}^{+0.4}$ & $(1.02_{-1.0}^{+1.0})\times 10^{-1}$ & $1.00$\\
$8.3\times10^{-2}-1.0\times10^{-1}$ & $(7.31_{-2.4}^{+3.2})\times 10^{-1}$ & $(7.31_{-2.4}^{+3.2})\times 10^{-1}$ &   & $1.00$\\
$1.0\times10^{-1}-1.3\times10^{-1}$ & $(5.18_{-1.9}^{+1.9})\times 10^{-1}$ & $(5.18_{-1.9}^{+1.9})\times 10^{-1}$ &   & $1.00$\\
$1.3\times10^{-1}-1.6\times10^{-1}$ & $(5.17_{-1.6}^{+1.6})\times 10^{-1}$ & $(5.17_{-1.6}^{+1.6})\times 10^{-1}$ &   & $1.00$\\
$1.6\times10^{-1}-2.1\times10^{-1}$ & $(1.65_{-0.8}^{+0.8})\times 10^{-1}$ & $(1.65_{-0.8}^{+0.8})\times 10^{-1}$ &   & $1.00$\\
$2.1\times10^{-1}-2.6\times10^{-1}$ & $(1.65_{-0.7}^{+0.7})\times 10^{-1}$ & $(1.65_{-0.7}^{+0.7})\times 10^{-1}$ &   & $1.00$\\
$2.6\times10^{-1}-3.2\times10^{-1}$ & $(5.25_{-5.3}^{+2.6})\times 10^{-2}$ & $(5.25_{-5.3}^{+2.6})\times 10^{-2}$ &   & $1.00$\\
$3.2\times10^{-1}-4.1\times10^{-1}$ & $(2.10_{-2.1}^{+2.1})\times 10^{-2}$ & $(2.10_{-2.1}^{+2.1})\times 10^{-2}$ &   & $1.00$\\
$5.1\times10^{-1}-6.4\times10^{-1}$ & $(2.67_{-2.7}^{+1.3})\times 10^{-2}$ & $(2.67_{-2.7}^{+1.3})\times 10^{-2}$ &   & $1.00$\\
$6.4\times10^{-1}-8.0\times10^{-1}$ & $(2.13_{-1.1}^{+1.1})\times 10^{-2}$ & $(2.13_{-1.1}^{+1.1})\times 10^{-2}$ &   & $1.00$\\
$1.0-1.3$ & $(6.78_{-6.8}^{+6.8})\times 10^{-3}$ & $(6.78_{-6.8}^{+6.8})\times 10^{-3}$ &   & $1.00$\\
\enddata
\end{deluxetable*}
 
\begin{deluxetable*}{llllc}
\tablecaption{The $150$~GHz differential counts.
\label{tab:diffcountstable150}}
\tablehead{
\colhead{Flux range} & \colhead{$dN/dS$ total} & \colhead{$dN/dS$ sync} & \colhead{$dN/dS$ dust} & \colhead{completeness} \\
\colhead{Jy} & \colhead{${\rm Jy}^{-1} {\rm deg}^{-2}$} & \colhead{${\rm Jy}^{-1} {\rm deg}^{-2}$} & \colhead{${\rm Jy}^{-1} {\rm deg}^{-2}$} & 
}
\startdata
$4.4\times10^{-3}-5.6\times10^{-3}$ & $(4.17_{-0.9}^{+0.9})\times 10^{2}$ & $(3.13_{-0.8}^{+0.8})\times 10^{2}$ & $(1.04_{-0.5}^{+0.5})\times 10^{2}$ & $0.89$\\
$5.6\times10^{-3}-7.0\times10^{-3}$ & $(2.31_{-0.2}^{+0.2})\times 10^{2}$ & $(1.89_{-0.2}^{+0.2})\times 10^{2}$ & $(4.11_{-1.0}^{+1.0})\times 10^{1}$ & $0.85$\\
$7.0\times10^{-3}-8.7\times10^{-3}$ & $(1.30_{-0.1}^{+0.1})\times 10^{2}$ & $(1.11_{-0.1}^{+0.1})\times 10^{2}$ & $(1.97_{-0.5}^{+0.5})\times 10^{1}$ & $0.97$\\
$8.7\times10^{-3}-1.1\times10^{-2}$ & $(7.17_{-0.8}^{+0.9})\times 10^{1}$ & $(6.35_{-0.8}^{+0.8})\times 10^{1}$ & $8.23_{-2.9}^{+2.9}$ & $1.00$\\
$1.1\times10^{-2}-1.4\times10^{-2}$ & $(4.17_{-0.6}^{+0.6})\times 10^{1}$ & $(3.84_{-0.5}^{+0.6})\times 10^{1}$ & $2.81_{-1.4}^{+1.9}$ & $1.00$\\
$1.4\times10^{-2}-1.7\times10^{-2}$ & $(2.84_{-0.4}^{+0.4})\times 10^{1}$ & $(2.73_{-0.4}^{+0.4})\times 10^{1}$ & $1.12_{-0.7}^{+0.7}$ & $1.00$\\
$1.7\times10^{-2}-2.2\times10^{-2}$ & $(1.76_{-0.2}^{+0.3})\times 10^{1}$ & $(1.73_{-0.3}^{+0.3})\times 10^{1}$ & $(2.98_{-3.0}^{+6.0})\times 10^{-1}$ & $1.00$\\
$2.2\times10^{-2}-2.7\times10^{-2}$ & $(1.31_{-0.2}^{+0.2})\times 10^{1}$ & $(1.26_{-0.2}^{+0.2})\times 10^{1}$ & $(4.76_{-2.4}^{+4.8})\times 10^{-1}$ & $1.00$\\
$2.7\times10^{-2}-3.4\times10^{-2}$ & $8.36_{-1.3}^{+1.5}$ & $8.17_{-1.3}^{+1.3}$ & $(1.90_{-1.9}^{+3.8})\times 10^{-1}$ & $1.00$\\
$3.4\times10^{-2}-4.2\times10^{-2}$ & $5.46_{-0.9}^{+1.1}$ & $5.46_{-0.9}^{+1.1}$ &   & $1.00$\\
$4.2\times10^{-2}-5.3\times10^{-2}$ & $3.02_{-0.6}^{+0.7}$ & $3.02_{-0.6}^{+0.7}$ &   & $1.00$\\
$5.3\times10^{-2}-6.7\times10^{-2}$ & $1.54_{-0.4}^{+0.5}$ & $1.54_{-0.4}^{+0.5}$ &   & $1.00$\\
$6.7\times10^{-2}-8.3\times10^{-2}$ & $(8.47_{-3.1}^{+2.3})\times 10^{-1}$ & $(8.47_{-3.1}^{+2.3})\times 10^{-1}$ &   & $1.00$\\
$8.3\times10^{-2}-1.0\times10^{-1}$ & $(9.22_{-2.5}^{+3.1})\times 10^{-1}$ & $(9.22_{-2.5}^{+3.1})\times 10^{-1}$ &   & $1.00$\\
$1.0\times10^{-1}-1.3\times10^{-1}$ & $(5.40_{-1.5}^{+2.0})\times 10^{-1}$ & $(5.40_{-1.5}^{+2.0})\times 10^{-1}$ &   & $1.00$\\
$1.3\times10^{-1}-1.6\times10^{-1}$ & $(2.35_{-1.2}^{+0.8})\times 10^{-1}$ & $(1.96_{-0.8}^{+0.8})\times 10^{-1}$ & $(3.91_{-3.9}^{+3.9})\times 10^{-2}$ & $1.00$\\
$1.6\times10^{-1}-2.1\times10^{-1}$ & $(1.56_{-0.6}^{+0.9})\times 10^{-1}$ & $(1.56_{-0.6}^{+0.9})\times 10^{-1}$ &   & $1.00$\\
$2.1\times10^{-1}-2.6\times10^{-1}$ & $(2.49_{-2.5}^{+2.5})\times 10^{-2}$ & $(2.49_{-2.5}^{+2.5})\times 10^{-2}$ &   & $1.00$\\
$2.6\times10^{-1}-3.2\times10^{-1}$ & $(7.95_{-4.0}^{+4.0})\times 10^{-2}$ & $(7.95_{-4.0}^{+4.0})\times 10^{-2}$ &   & $1.00$\\
$3.2\times10^{-1}-4.1\times10^{-1}$ & $(4.76_{-3.2}^{+1.6})\times 10^{-2}$ & $(4.76_{-3.2}^{+1.6})\times 10^{-2}$ &   & $1.00$\\
$4.1\times10^{-1}-5.1\times10^{-1}$ & $(2.53_{-1.3}^{+1.3})\times 10^{-2}$ & $(2.53_{-1.3}^{+1.3})\times 10^{-2}$ &   & $1.00$\\
$5.1\times10^{-1}-6.4\times10^{-1}$ & $(2.02_{-1.0}^{+2.0})\times 10^{-2}$ & $(2.02_{-1.0}^{+2.0})\times 10^{-2}$ &   & $1.00$\\
$6.4\times10^{-1}-8.0\times10^{-1}$ & $ 0 _{- 0 }^{+1.6\times10^{-2}}$ & $ 0 _{- 0 }^{+1.6\times10^{-2}}$ &   & $1.00$\\
$8.0\times10^{-1}-1.0$ & $(6.43_{-6.4}^{+6.4})\times 10^{-3}$ & $(6.43_{-6.4}^{+6.4})\times 10^{-3}$ &   & $1.00$\\
$1.0-1.3$ & $(5.13_{-5.1}^{+5.1})\times 10^{-3}$ & $(5.13_{-5.1}^{+5.1})\times 10^{-3}$ &   & $1.00$\\
\enddata
\end{deluxetable*}
 
\begin{deluxetable*}{llllc}
\tablecaption{The $220$~GHz differential counts.
\label{tab:diffcountstable220}}
\tablehead{
\colhead{Flux range} & \colhead{$dN/dS$ total} & \colhead{$dN/dS$ sync} & \colhead{$dN/dS$ dust} & \colhead{completeness} \\
\colhead{Jy} & \colhead{${\rm Jy}^{-1} {\rm deg}^{-2}$} & \colhead{${\rm Jy}^{-1} {\rm deg}^{-2}$} & \colhead{${\rm Jy}^{-1} {\rm deg}^{-2}$} & 
}
\startdata
$1.1\times10^{-2}-1.4\times10^{-2}$ & $(6.93_{-2.5}^{+3.0})\times 10^{1}$ & $(2.48_{-1.5}^{+2.0})\times 10^{1}$ & $(4.46_{-2.0}^{+2.0})\times 10^{1}$ & $0.84$\\
$1.4\times10^{-2}-1.7\times10^{-2}$ & $(4.42_{-1.7}^{+1.7})\times 10^{1}$ & $(2.04_{-1.0}^{+1.0})\times 10^{1}$ & $(2.38_{-1.4}^{+1.4})\times 10^{1}$ & $0.98$\\
$1.7\times10^{-2}-2.2\times10^{-2}$ & $(2.59_{-0.6}^{+0.9})\times 10^{1}$ & $(1.48_{-0.5}^{+0.6})\times 10^{1}$ & $(1.11_{-0.5}^{+0.5})\times 10^{1}$ & $1.00$\\
$2.2\times10^{-2}-2.7\times10^{-2}$ & $(1.63_{-0.3}^{+0.3})\times 10^{1}$ & $(1.07_{-0.2}^{+0.2})\times 10^{1}$ & $5.61_{-1.5}^{+1.5}$ & $0.93$\\
$2.7\times10^{-2}-3.4\times10^{-2}$ & $8.95_{-1.5}^{+1.7}$ & $6.67_{-1.5}^{+1.3}$ & $2.29_{-0.8}^{+1.0}$ & $1.00$\\
$3.4\times10^{-2}-4.2\times10^{-2}$ & $4.55_{-1.1}^{+1.1}$ & $3.34_{-0.9}^{+0.9}$ & $1.21_{-0.6}^{+0.5}$ & $1.00$\\
$4.2\times10^{-2}-5.3\times10^{-2}$ & $2.78_{-0.6}^{+0.7}$ & $2.42_{-0.6}^{+0.7}$ & $(3.63_{-2.4}^{+2.4})\times 10^{-1}$ & $1.00$\\
$5.3\times10^{-2}-6.7\times10^{-2}$ & $1.26_{-0.4}^{+0.5}$ & $1.16_{-0.4}^{+0.5}$ & $(9.65_{-9.7}^{+9.7})\times 10^{-2}$ & $1.00$\\
$6.7\times10^{-2}-8.3\times10^{-2}$ & $1.23_{-0.3}^{+0.4}$ & $1.16_{-0.3}^{+0.3}$ & $(7.70_{-7.7}^{+7.7})\times 10^{-2}$ & $1.00$\\
$8.3\times10^{-2}-1.0\times10^{-1}$ & $(6.76_{-2.5}^{+2.5})\times 10^{-1}$ & $(5.53_{-1.8}^{+2.5})\times 10^{-1}$ & $(1.23_{-0.6}^{+0.6})\times 10^{-1}$ & $1.00$\\
$1.0\times10^{-1}-1.3\times10^{-1}$ & $(2.45_{-1.0}^{+1.5})\times 10^{-1}$ & $(1.96_{-1.0}^{+1.5})\times 10^{-1}$ & $(4.90_{-4.9}^{+4.9})\times 10^{-2}$ & $1.00$\\
$1.3\times10^{-1}-1.6\times10^{-1}$ & $(1.96_{-0.8}^{+1.2})\times 10^{-1}$ & $(1.96_{-0.8}^{+1.2})\times 10^{-1}$ &   & $1.00$\\
$1.6\times10^{-1}-2.1\times10^{-1}$ & $(9.37_{-6.2}^{+6.2})\times 10^{-2}$ & $(9.37_{-6.2}^{+6.2})\times 10^{-2}$ &   & $1.00$\\
$2.1\times10^{-1}-2.6\times10^{-1}$ & $(4.98_{-2.5}^{+5.0})\times 10^{-2}$ & $(4.98_{-2.5}^{+5.0})\times 10^{-2}$ &   & $1.00$\\
$2.6\times10^{-1}-3.2\times10^{-1}$ & $(5.96_{-4.0}^{+4.0})\times 10^{-2}$ & $(3.98_{-2.0}^{+4.0})\times 10^{-2}$ & $(1.99_{-2.0}^{+2.0})\times 10^{-2}$ & $1.00$\\
$3.2\times10^{-1}-4.1\times10^{-1}$ & $(3.17_{-1.6}^{+3.2})\times 10^{-2}$ & $(3.17_{-1.6}^{+3.2})\times 10^{-2}$ &   & $1.00$\\
$4.1\times10^{-1}-5.1\times10^{-1}$ & $(3.80_{-2.5}^{+2.5})\times 10^{-2}$ & $(3.80_{-2.5}^{+2.5})\times 10^{-2}$ &   & $1.00$\\
$5.1\times10^{-1}-6.4\times10^{-1}$ & $ 0 _{- 0 }^{+2.0\times10^{-2}}$ & $ 0 _{- 0 }^{+2.0\times10^{-2}}$ &   & $1.00$\\
$6.4\times10^{-1}-8.0\times10^{-1}$ & $(8.06_{-8.1}^{+8.1})\times 10^{-3}$ & $(8.06_{-8.1}^{+8.1})\times 10^{-3}$ &   & $1.00$\\
$8.0\times10^{-1}-1.0$ & $ 0 _{- 0 }^{+1.3\times10^{-2}}$ & $ 0 _{- 0 }^{+1.3\times10^{-2}}$ &   & $1.00$\\
$1.0-1.3$ & $(5.13_{-5.1}^{+5.1})\times 10^{-3}$ & $(5.13_{-5.1}^{+5.1})\times 10^{-3}$ &   & $1.00$\\
\enddata
\end{deluxetable*}


\section{Number counts}
\label{sec:number_counts}

We derive source number counts using a bootstrap method outlined in \citet{austermann09}. For each source in the catalog, we randomly draw 50,000 fluxes from the deboosted three-band flux posterior, $P(S_{95},S_{150},S_{220})$. We thus obtain 50,000 mock source catalogs. We resample each of those mock catalogs by drawing with replacement a number of sources that is a Poisson deviate of the catalog size. For each of the resampled catalogs, we compute the number counts $dN/dS$ in each flux bin. We correct the counts for completeness in each bin based on the simulations described in  \S\ref{sec:completeness}. We perform this procedure separately for each field.

We do not explicitly correct for purity, as it is intrinsically accounted for in the Bayesian deboosting as follows. Some sources in the mock catalogs will be assigned sub-threshold fluxes due to drawing from the region of the flux posterior that is below the detection threshold and will thus be thrown out of the counts.

We combine the number counts from different fields by summing up the counts, weighted by a quantity we denote as ``effective area''. We define this as the area of the field multiplied by the completeness in each flux bin. We then use the cumulative distributions of $dN/dS$ over all catalogs to obtain the $16\%$,  $50\%$, and  $84\%$ percentile points, which represent the median and equivalent 1$\sigma$ errors on the final counts. 
Because the fields have varying depths, the lowest few flux bins only contain contributions from fields with detection thresholds below the bin range. 
We use all five fields in Table~\ref{tab:fields} in the number counts.
Thus, the 95 GHz counts reflect the three 2009 fields, or 584 \sqdeg, while the 150 and 220 GHz counts reflect all five fields, totalling 771 \sqdeg.

\begin{figure*}[!t]
\begin{center}
  \subfigure{\includegraphics[width=8.9cm]{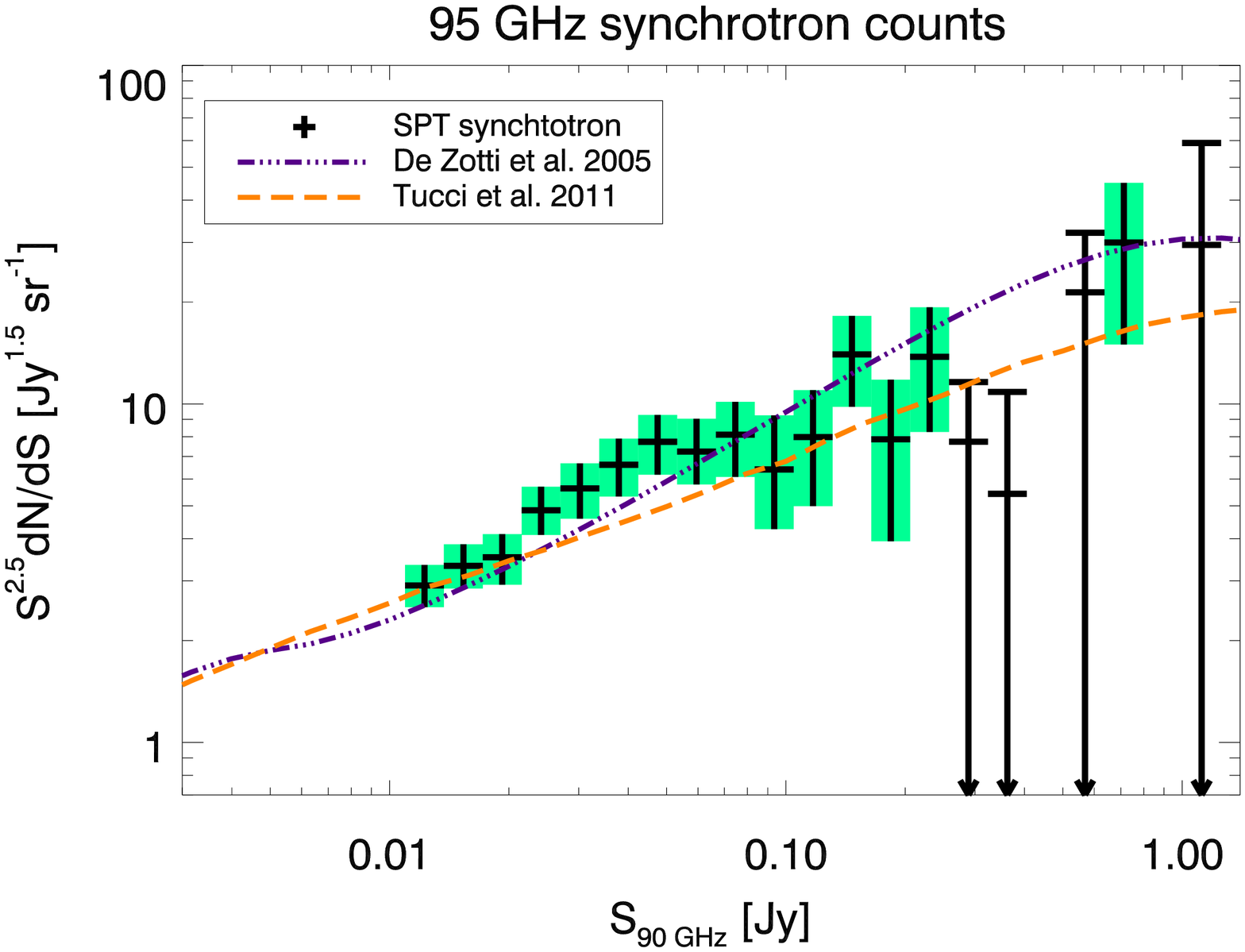}}
  \subfigure{\includegraphics[width=8.9cm]{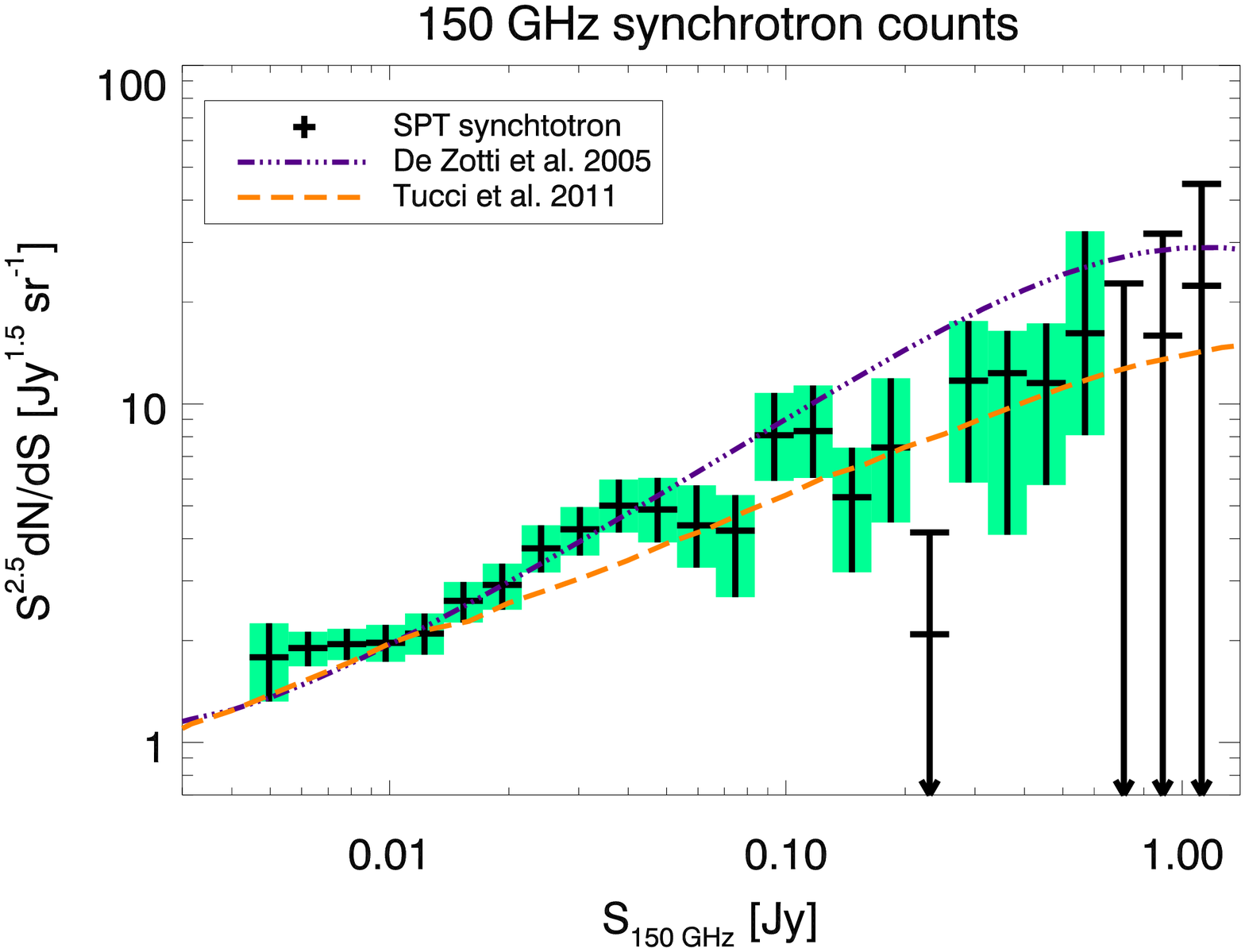}}
  \subfigure{\includegraphics[width=8.9cm]{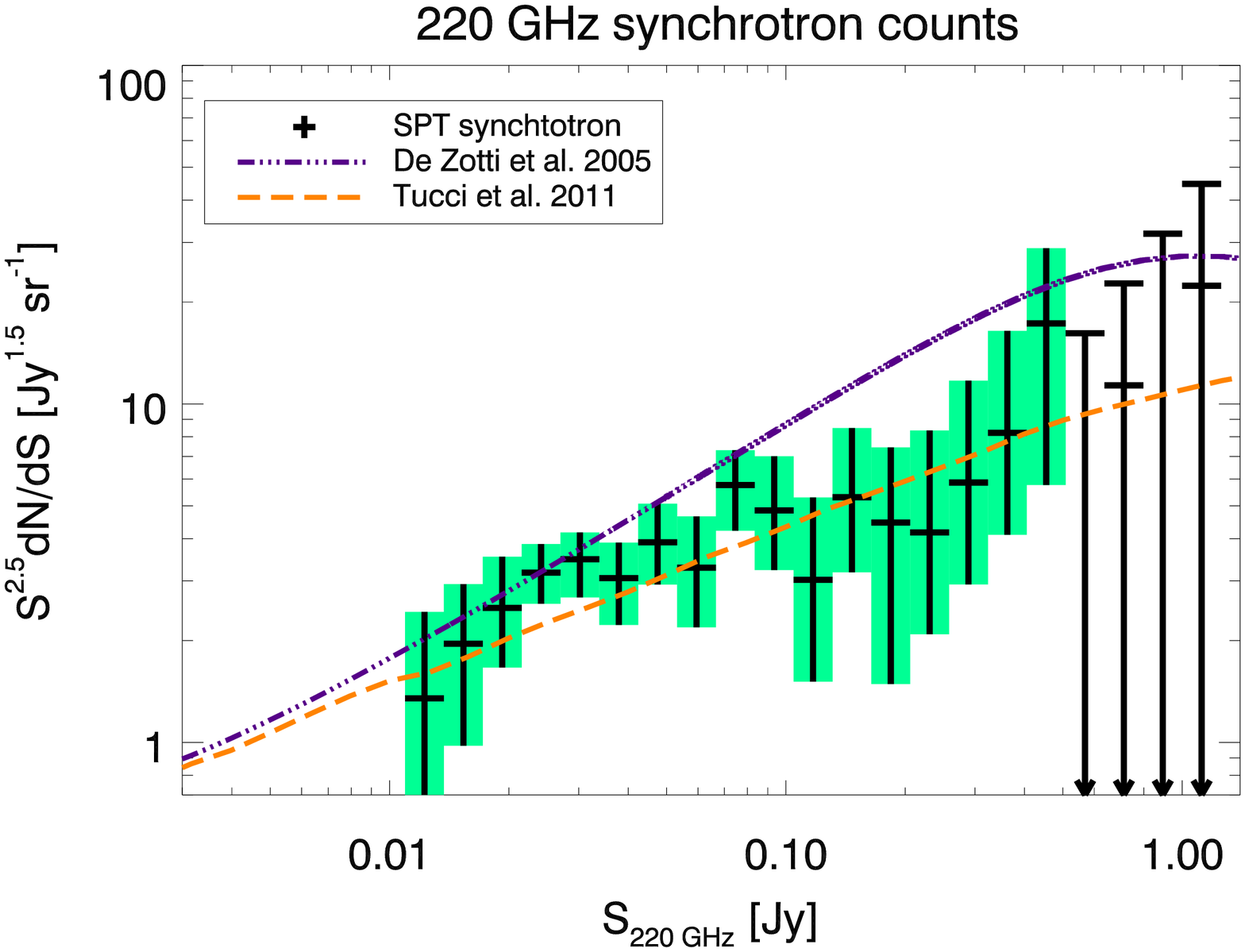}}
\end{center}
\caption{Number counts of SPT synchrotron-dominated sources. Overplotted are the \citet{dezotti05} and the \citet{tucci11} models.
\label{fig:sync_counts}}
\end{figure*}

We account for sources of uncertainty as follows. Taking Poisson deviates of the real catalog size for the mocks accounts for sample variance. We do not include the uncertainty from variance due to large scale structure, as the large survey area assures sufficient sampling of structure in the universe.
As mentioned in \S\ref{sec:deboost}, because the beam and calibration error are the same for all sources in the catalog, we use a set of flux posteriors constructed without including the beam and calibration error in the covariance matrix for each source. Rather, we incorporate a realization of beam and calibration noise that is common to all sources in a mock catalog, but is different between catalogs. The source flux posterior includes errors due to map noise and cross-band deboosting. We note that the errors on the number counts are correlated between bins, roughly at the 5$\%$ level.

Extended sources contribute less than 8$\%$ of the counts in any flux  bin, and typically less than 3$\%$; the effect of their underestimated fluxes is completely subdominant to the statistical errors on the number counts.

We derive number counts for the two source populations using a probabilistic classification method. For each source in the resampled catalogs, which stands as a triplet of fluxes drawn from a posterior, we calculate the spectral index $\alphat$ and classify the source as dusty if it exceeds the threshold index. It follows that a source which has $P(\alpha \ge 1.5) = p$ will be included in the dusty counts in a fraction $p$ of the resamplings and in the synchrotron counts in the remaining $1-p$ fraction.

Figure~\ref{fig:counts} shows source number counts in the three frequency bands. We show the total counts, as well as counts for the synchrotron- and dust-dominated populations. Synchrotron sources are the main component everywhere except for the lowest flux bins at 220 GHz, where the dust component becomes dominant. 
The counts are consistent with the results published in V10. 

\begin{deluxetable*}{ l | c c c |c c c| c c c}
\tablecaption{AGN model goodness of fit} \small

\tablehead{
& & 95 GHz & & & 150 GHz & & & 220 GHz & \\
Model & $\chi^2$ & DOF & PTE & $\chi^2$ & DOF & PTE & $\chi^2$ & DOF & PTE }
\startdata

De Zotti et al. (2005)  &  132.717 &  21  & 0  & 78.671  & 25 & $1.834 \times 10^{-7}$  & 86.751 & 21 &  0 \\
Tucci et al. (2011)     &  54.704  & 21   & 0  & 31.386  & 25 & 0.177  & 12.587 & 21 & 0.922

\enddata

\label{tab:agn_fit}
\tablecomments{Goodness of fit for the synchrotron number counts models. We list the $\chi^2$ value between the data and the models, the number of degrees of freedom (DOF) for the fit and the probability to exceed (PTE) the $\chi^2$ value.}
\end{deluxetable*}

Figure~\ref{fig:comparison} presents a comparison between our 150~GHz counts, 143~GHz number counts from Planck \citep{planck12-7}, and 148~GHz number counts from ACT \citep{marsden13} (left panel); and our 220~GHz counts, Planck 217~GHz counts, and ACT 218~GHz counts (right panel). The three sets of counts are consistent with one another.

\section{Discussion}
\label{sec:pop}

\subsection{Source populations}

In \S\ref{sec:alpha}, we classify sources based on their posterior $\alphat$ probability distribution. 
However, given the three observing bands, the picture is inevitably more complicated. 
Apart from purely falling or rising spectra, we also see spectra that seem to dip or peak within our frequency range. 
We stress that we define ``dipping'' and ``peaking'' sources by the criteria listed in Table \ref{tab:spectral}, such that there might not be an actual trough or peak in the spectrum.

We note that it is possible for sources to scatter out of the standard ``falling'' and ``rising'' quadrants into the ``peaking" or ``dipping" quadrants, especially at low significance. Considering a 7$\sigma$ detection at 150 GHz (roughly the average significance in the 6-12 mJy column in Table \ref{tab:spectral}), a source with a true $\alphao=\alphat=-0.7$ gets misclassified as ``peaking'' 1\% of the time, and a source with a true $\alphao=\alphat=3.5$ gets misclassified as ``dipping'' 1\% of the time.

The first category (``dipping" sources) comprises what appears to be a synchrotron source at 95 GHz, with dust emission picking up between 150 and 220 GHz. Such sources 
are expected to be low-redshift ULIRGs or regular spirals. For instance, a source that is spectrally similar to Arp 220, a typical ULIRG \citep{silva98}, but had slightly more dust emission,  would appear in this quadrant of spectral index parameter space. 
The brightest ``dipping'' sources are nearby galaxies from the NGC catalog that show both strong radio and starburst activity. Many of these galaxies have counterparts in SUMSS or IRAS.

The ``peaking'' sources are the least numerous population, and the detections are lower signal-to-noise, so the spectral indices are more uncertain, but we might see evidence of a self-absorbed synchrotron component in the SED. 
This is believed to be the emission mechanism in the case of Gigahertz-Peaked Spectrum sources \citep{odea98}, although the subclass with higher turnover frequencies, High Frequency Peakers, still typically peaks at tens of GHz \citep{dallacasa00}. 
Most of the brightest such galaxies have radio SUMSS counterparts. 
Two interesting cases to note are the second and sixth brightest peaking sources, which seem to be associated with the pulsating stars X Pav and NU Pav. 
About half of the ``peaking'' sources do not have counterparts in the external catalogs that we have checked.

\subsection{Number counts by source population}

In this subsection, we will consider models of galaxy number counts from the literature and compare them to the measured counts.

\subsubsection{Synchtrotron-dominated sources}

\begin{figure*}
\begin{center}
  \subfigure{\includegraphics[width=8.9cm]{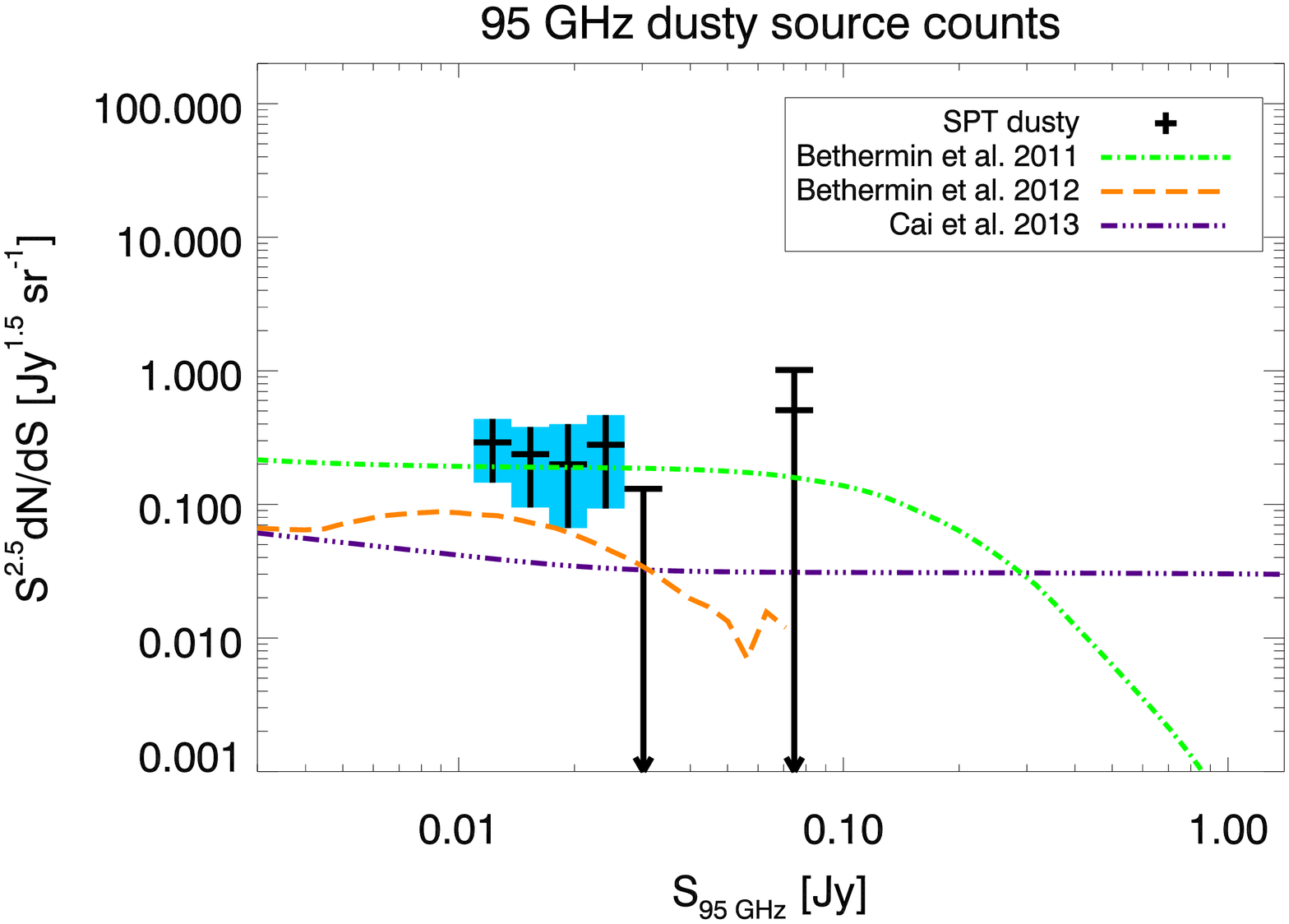}}
  \subfigure{\includegraphics[width=8.9cm]{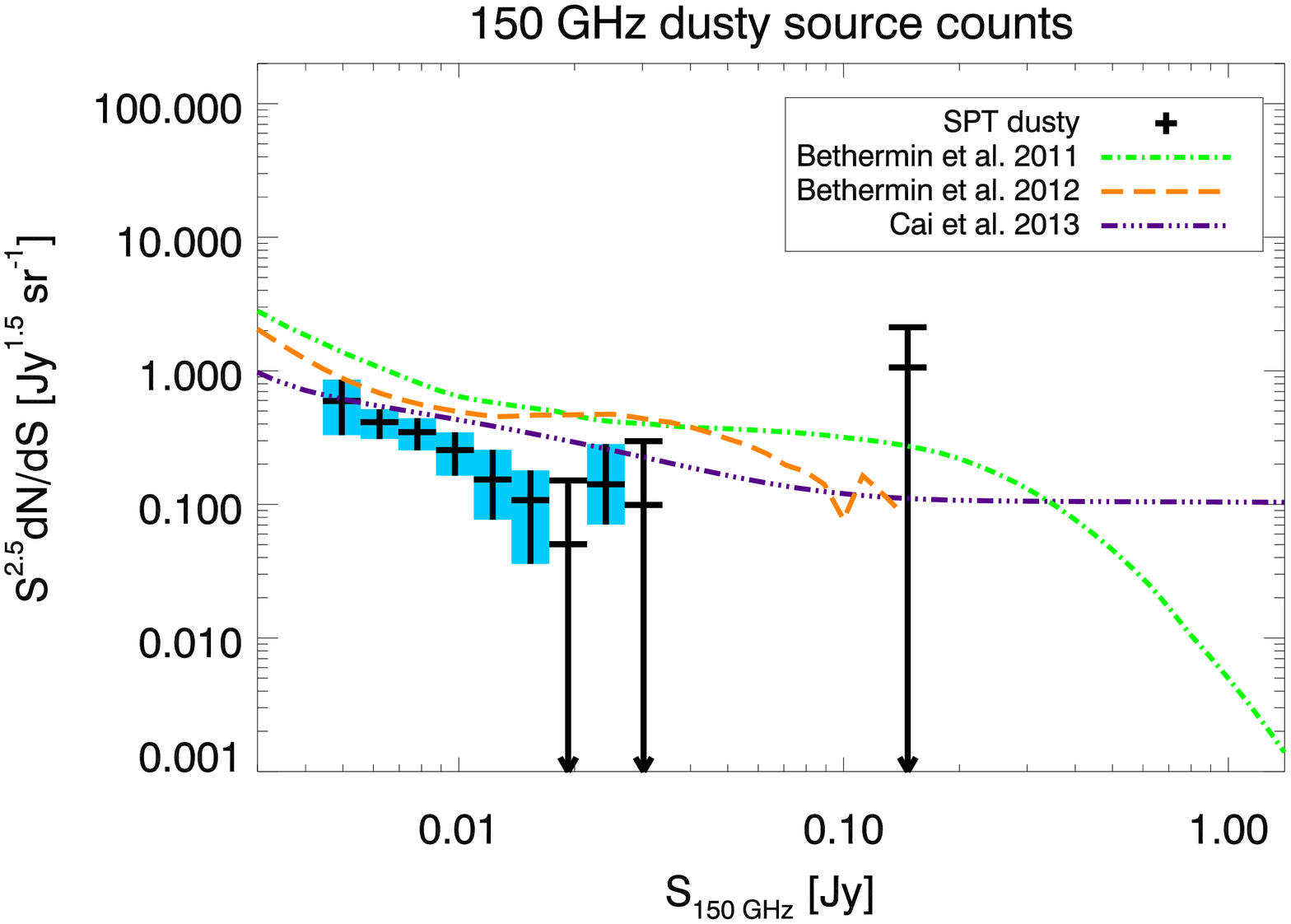}}
  \subfigure{\includegraphics[width=8.9cm]{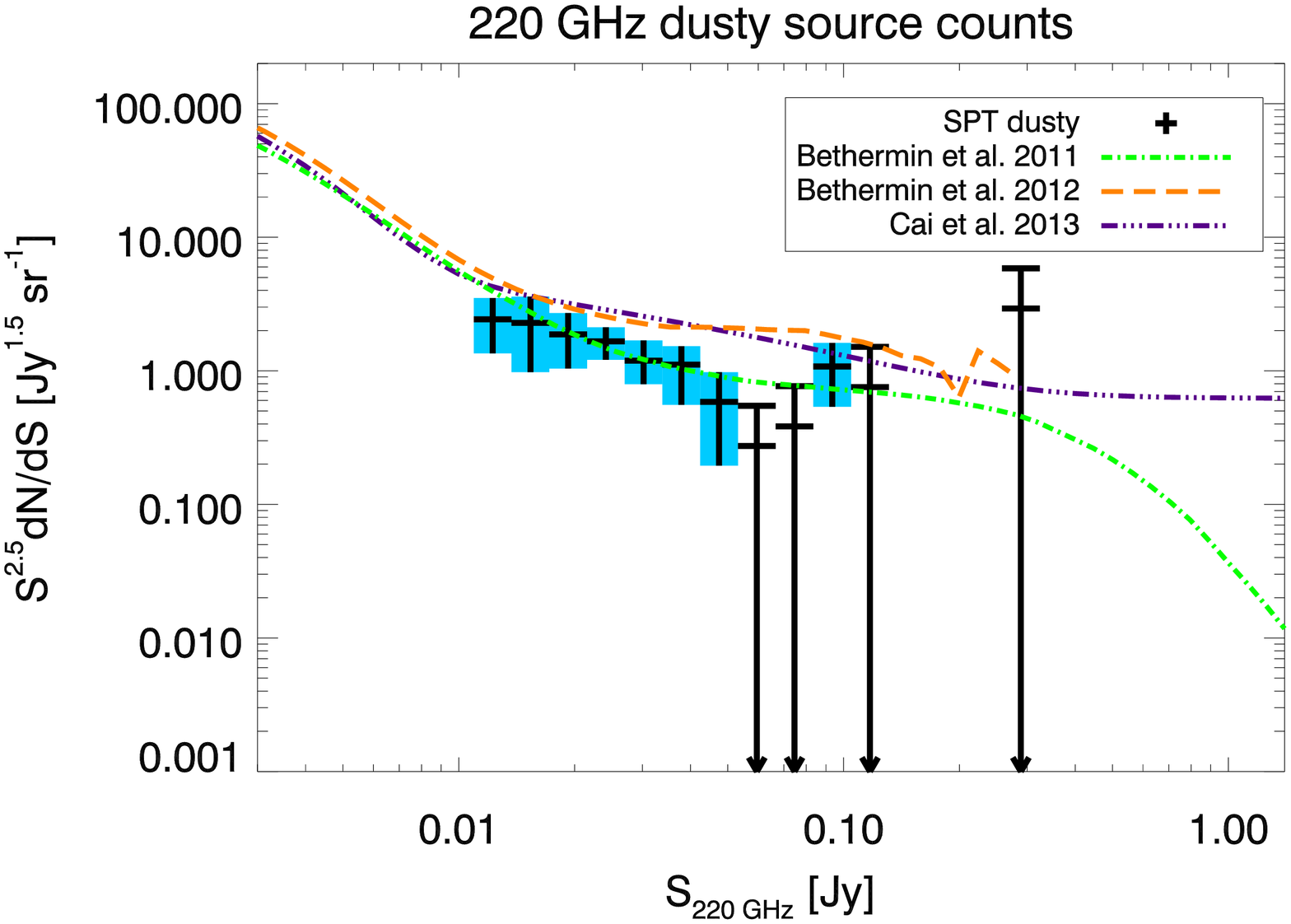}}
\end{center}
\caption{Number counts of SPT dust-dominated sources. Overplotted are the \citet{bethermin11}, \citet{bethermin12}, and \citet{cai13} models.
\label{fig:dust_counts}}
\end{figure*}

\begin{deluxetable*}{ l | c c c |c c c| c c c}
\tablecaption{Dusty model goodness of fit} \small

\tablehead{
& & 95 GHz & & & 150 GHz & & & 220 GHz & \\
Model & $\chi^2$ & DOF & PTE & $\chi^2$ & DOF & PTE & $\chi^2$ & DOF & PTE 
}
\startdata

Bethermin et al. (2011)     & 3.839  & 6   & 0.700  & 231.192  & 10 & 0  & 7.464 & 12 & 0.825 \\
Bethermin et al. (2012) & 7.086 & 6 & 0.313 & 123.991 & 10 & 0 & 94.894 & 12 & 0 \\
Cai et al. (2013)  &  9.292 &  6  & 0.158  & 44.765  & 10 & 0  & 81.117 & 12 & 0

\enddata

\label{tab:dust_fit}
\tablecomments{Goodness of fit for the dusty number counts models. We list the $\chi^2$ value between the data and the models, the number of degrees of freedom (DOF) for the fit and the probability to exceed (PTE) the $\chi^2$ value.}
\end{deluxetable*}

\begin{figure}
\includegraphics[width=8.9cm]{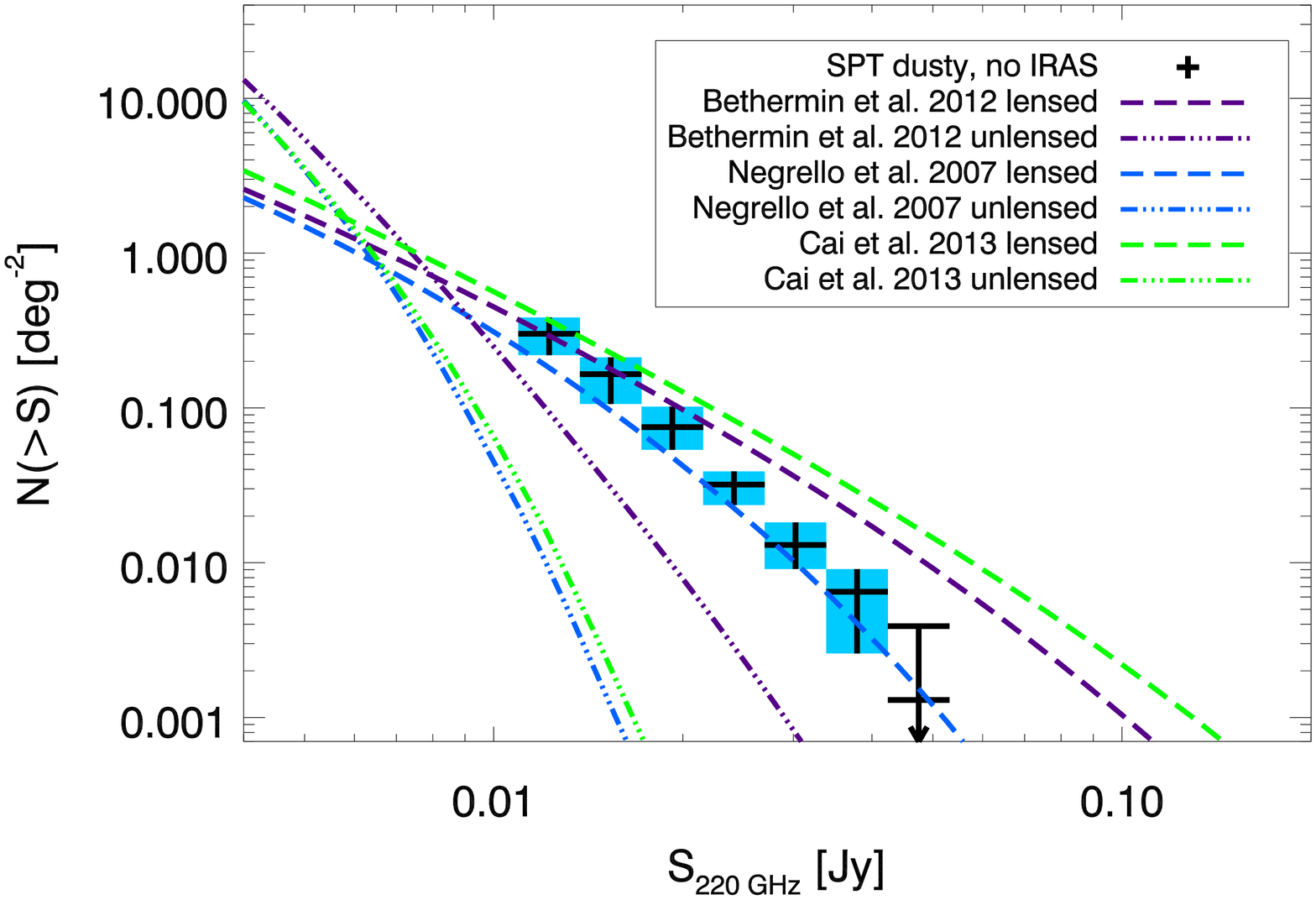}
\caption{Number counts of SPT dust-dominated sources excluding sources with counterparts in the IRAS catalog. Overplotted are the lensed components of the \citet{bethermin12}, Negrello et al. (2007), and \citet{cai13} models.
\label{fig:dust_ngts}}
\end{figure}

Figure~\ref{fig:sync_counts} shows number counts for the synchrotron-dominated population, plotted against the \citet{dezotti05} and \citet{tucci11} models.

The \citet{dezotti05} model takes into account flat- and steep-spectrum radio sources, where the steep-spectrum category includes dusty spheroidals and GHz peaked spectrum sources. The model extrapolates blazar spectra using a simple power-law approximation with a spectral index $\alpha\simeq -0.1$ above 100 GHz.

The \citet{tucci11} model is constructed based on extrapolations of number counts from high radio frequencies (5 GHz). It considers the spectral behavior of the different source populations, flat-spectrum (FSRQs and BL Lac), steep-spectrum and inverted spectrum, in a statistical way and takes into account the main physical mechanisms responsible for the emission. The model features different distributions of spectral break frequencies for FSRQs and BL Lacs. We compare our counts to the ``C2Ex'' version of this model, which was found by the authors to best fit available high-frequency ($\nu>$100 GHz) counts.

Table \ref{tab:agn_fit} lists the $\chisq$ values for the synchrotron-dominated model comparisons. The \citet{dezotti05} model fits the lower flux range rather well and is also a good fit in the intermediate range at 150 GHz, while slightly underpredicting intermediate 95 GHz counts. However, the model is in excess of the data at the high-flux end in all frequency bands. This behavior is most likely due to the simple power-law extrapolation that the model is based on; neglecting the presence of a spectral break leads to overpredicting the number of bright blazars at these frequencies.

The \citet{tucci11} model improves upon the former by incorporating the effects of spectral steepening. Consequently, the model is a good fit to our data above ~80 mJy and below ~20 mJy in all bands, but underpredicts the counts in the intermediate flux range at 95 and 150 GHz. In the 220 GHz band, except for a few bins, the \citet{tucci11} model comes very close to our counts.

\subsubsection{Dust-dominated sources}

Figure~\ref{fig:dust_counts} shows number counts for dust-dominated sources. Overplotted are the \citet{bethermin11}, \citet{bethermin12}, and \citet{cai13} models.

The \citet{bethermin11} model is a parametric backwards evolution model which considers normal and starburst galaxies and is based on an evolution in density and luminosity of the luminosity function, tuned to reproduce a large set of observational constraints---although none of the observational constraints are at SPT observing frequencies. This model includes a strong lensing contribution from high-redshift SMGs. 

\citet{bethermin12} is an empirical model based on two star formation modes, corresponding to main sequence and starburst galaxies. It considers the redshift evolution of these two populations and incorporates two corresponding families of SEDs derived from Herschel observations. This model includes the effect of strong lensing on the counts as well, using the lensing prescription of \citet{hezaveh11}. All parameters are constrained by non-SPT observations and have not been tuned to fit the SPT counts.

\citet{cai13} combine a physical forward model for spheroidal galaxies and the early evolution of the associated AGN with a phenomenological backward model for late-type galaxies and for the later AGN evolution. It is calibrated using data from mid-infrared to millimeter wavelengths.

Table \ref{tab:dust_fit} lists the $\chisq$ values for the dusty model comparisons. The \citet{bethermin11} model is a very good fit to the data at 95 and 220 GHz, but overpredicts the counts at 150 GHz. It is not clear what causes this behavior.

The \citet{bethermin12} model underpredicts the 95 GHz counts, overpredicts the 150 GHz counts above 10 mJy, and overpredicts the 220 GHz counts. This suggests that the model might be assuming too steep a slope for the SED between 95 and 220 GHz. This is plausible since the SED library was calibrated from the far infrared down to 1.1 mm ($\sim$270 GHz) and extrapolated down to lower frequencies. Slightly warmer local templates would bring down the 150 and 220 GHz counts, while an increase in the synchrotron and/or free-free emission would boost the 95 GHz counts, bringing the model in agreement with the data. The drop in counts at very bright flux for both the \citet{bethermin11} and the \citet{bethermin12} models is an artifact of the redshift grid; they should, in fact, converge to a flat behavior. 

The \citet{cai13} model underpredicts the 95 GHz counts, overpredicts the mid-flux range 150 GHz counts, and overpredicts the 220 GHz counts.

Figure~\ref{fig:dust_ngts} shows the dust-dominated SPT number counts, excluding sources that have counterparts in the IRAS catalog. We overplot the lensed and unlensed components of the \citet{bethermin12}, \citet{negrello07}, and \citet{cai13} models. To mimic the IRAS exclusion, sources with 60 $\mu$m flux greater than 200 mJy have been removed from the \citet{negrello07} model. We have excluded sources below a redshift of 0.5 from the \citet{bethermin12} model and also removed low-redshift populations from the \citet{cai13} model. The counts clearly exceed all the unlensed models and are better fitted by the lensed population models. In particular, the \citet{negrello07} model is an excellent fit to the data, suggesting that these counts are well explained by a lensed population of high-redshift dusty sources. The other two models agree at low flux but overpredict the counts at intermediate to high flux levels.

\subsubsection{Comparison to source model constraints from fluctuation measurements}

The number counts presented here probe the relatively high-flux end of mm-wave source populations, 
and these results are in some tension with published source count models. It is possible to probe to lower
fluxes using measurements of the uncorrelated (``Poisson'') point-source contribution to the fluctuation
power in the same maps that we use here to search for detectable sources. It is reasonable to ask whether
measurements of fluctuation power are also in tension with models, and, if so, if the tension would be 
alleviated with the same modifications to the models preferred by the source count data.
Recent studies of this fluctuation
power using SPT data include measurements of the Poisson point-source Fourier-domain
two-point function, or power spectrum, in \citet{reichardt12b} and Fourier-domain three-point function, 
or bispectrum, in \citet{crawford13}. Both of these works exclude sources detected above $5\sigma$ 
from the fluctuation analysis, so the results are almost fully independent of the source count 
results presented here.

We defer a detailed comparison of all three statistics (source 
counts, power spectrum, bispectrum) and all possible combinations of models to a future work; 
here we simply note that \citet{crawford13}, considering a subset of source count models 
we use in this work, found that no combination of models provided a statistically acceptable
fit to the Poisson point-source bispectrum in all three SPT bands. Specifically, both the \citet{dezotti10}
and \citet{tucci11} models underpredicted the 95 GHz bispectrum (which is expected to be dominated
by radio sources), and both the \citet{bethermin11} and \citet{bethermin12} models overpredicted the 
220 GHz bispectrum (which is expected to be dominated by DSFGs). At 150 GHz, where the 
radio and DSFG contributions to the bispectrum are both expected to be significant, the 
\citet{bethermin11} and \citet{bethermin12} predict significantly higher bispectrum levels than 
observed, such that the combined prediction is high regardless of which radio model is used.

If we break the total flux range probed by bispectrum and source counts into ``low-flux'' (below the 
detection threshold used in this work), ``moderate-flux'' (roughly 10 to 100 mJy), and ``high-flux'' 
(above 100 mJy) regimes, we can draw general conclusions about the radio and DSFG models
in the three regimes. We find that both radio source models appear to underpredict the measured 
counts in the low-flux regime at 95 GHz. The \citet{dezotti10} model agrees reasonably well in the 
moderate-flux regime but overpredicts the counts in the high-flux regime, while the \citet{tucci11} 
model underpredicts the moderate-flux counts but accurately predicts the high-flux counts. The 
DSFG source models considered here and in \citet{crawford13} appear to overpredict the 150 GHz counts at all flux levels.
The \citet{bethermin11} model accurately predicts the 220 GHz counts in the moderate- and high-flux
regimes but overpredicts the low-flux counts, while the \citet{bethermin12} model appears to 
overpredict the 220 GHz counts in all three regimes.

\section{Conclusion}
\label{sec:concl}

We have presented a 3-band catalog of 1545 sources from 771 \sqdeg~of the SPT-SZ survey. We have derived deboosted fluxes and spectral indices and have classified the sources into synchrotron- or dust-dominated populations based on their $\alpha^{150}_{220}$ spectral indices.

We have discovered a significant fraction of both synchrotron and dusty sources that have no counterparts in external catalogs. 
The dusty sources without counterparts represent our lensed SMG candidates: the SPT continues to discover sources that are likely to be high-redshift, strongly lensed submillimeter galaxies, providing interesting targets to follow up in the submillimeter and other wavebands, particularly with ALMA.
The synchrotron-dominated sources with no counterparts could be due simply to source variability, though the number of such sources in this catalog is in some statistical tension with the same number from earlier SPT results based on 87 \sqdeg~\citep{vieira10}.
We have also separated sources into four categories based on their position in the two-dimensional spectral index space. 

We have derived source number counts in the three SPT frequency bands, including total number counts and number counts for each of the two source populations. Synchrotron sources dominate the number counts everywhere except below 17 mJy in the 220 GHz band. 
The measured counts can be used to estimate levels of point-source foreground power for CMB 
analyses from ground-based CMB experiments or the Planck satellite.

We have also compared our measured counts to source count models for each population in each frequency band. We find small but significant discrepancies between our measured counts and all the models we consider for either population. 
The new information provided by our counts thus has the potential to inform models of galaxy formation and evolution as well as models of AGN behavior.

Work is ongoing to extend the analysis presented here to 
the full 2500 \sqdeg~SPT-SZ survey. 
This final catalog will include many new examples of high-redshift, strongly lensed dusty galaxies, and the number counts derived from the full catalog will have smaller uncertainties (by roughly a factor of $\sqrt{3}$), which will lead to improved constraints on models of galaxy formation and evolution, on AGN models, and on the contamination of CMB measurements by point sources.

\begin{acknowledgments}
The SPT is supported by the National Science Foundation through grant ANT-0638937, with partial support provided by NSF grant PHY-1125897, the Kavli Foundation, and the Gordon and Betty Moore Foundation. M.~Aravena was co-funded under the Marie Curie Actions of the European Commission (FP7-COFUND). R.~Keisler acknowledges support from NASA Hubble Fellowship grant HF-51275. 

This publication makes use of data products from sthe Wide-field Infrared Survey Explorer, which is a joint project of the University of California, Los Angeles, and the Jet Propulsion Laboratory/California Institute of Technology, funded by the National Aeronautics and Space Administration. This research uses observations from the AKARI mission, a JAXA project with the participation of ESA.

\end{acknowledgments}


\pagestyle{plain}

\end{document}